\magnification=1200
\hoffset=.0cm
\voffset=.0cm
\baselineskip=.53cm plus .53mm minus .53mm

%
%
\def\ref#1{\lbrack#1\rbrack}
%
%
%
%
\input amssym.def
\input amssym.tex
%
%
\font\teneusm=eusm10                    
\font\seveneusm=eusm7                   
\font\fiveeusm=eusm5                 
%
%

%
%
\font\cps=cmcsc10
%
%
\newfam\eusmfam
\textfont\eusmfam=\teneusm
\scriptfont\eusmfam=\seveneusm
\scriptscriptfont\eusmfam=\fiveeusm

\def\proclaim #1. #2\par{\medbreak{\cps #1.\enspace}{\it #2}\par\medbreak}
%
%
%
%

\def\real{{\rm Re}\hskip 1pt}
\def\imag{{\rm Im}\hskip 1pt}

\def\ker{{\rm ker}\hskip 1pt}

\def\ind{{\rm ind}\hskip 1pt}

\def\Aut{{\rm Aut}\hskip 1pt}

\def\SU{{\rm SU}\hskip 1pt}
\def\SO{{\rm SO}\hskip 1pt}
\def\Spin{{\rm Spin}\hskip 1pt}
\def\ee{{\rm e}}
\def\Sp{{\rm Sp}\hskip 1pt}
\def\GL{{\rm GL}\hskip 1pt}

\def\PGL{{\rm PGL}\hskip 1pt}
\def\hst1{\hskip 1pt}
\def\square{\,\vbox{\hrule \hbox{\vrule height 0.22 cm 
\hskip 0.22 cm \vrule height 0.22 cm}\hrule}\,}
%
%
%
%

\hbox to 16.5 truecm{July 1997   \hfil DFUB 97--9}
\hbox to 16.5 truecm{Version 1  \hfil gr-qc/9707048}
\vskip2cm
\centerline{\bf THE QUATERNIONIC GEOMETRY OF}   
\centerline{\bf 4D CONFORMAL FIELD THEORY}
\vskip1cm
\centerline{by}
\vskip.5cm
\centerline{\bf Roberto Zucchini}
\centerline{\it Dipartimento di Fisica, Universit\`a degli Studi di Bologna}
\centerline{\it V. Irnerio 46, I-40126 Bologna, Italy}
\vskip1cm
\centerline{\bf Abstract} 
We show that 4--dimensional conformal field theory is most naturally 
formulated on Kulkarni 4--folds, i. e. real 4--folds endowed with 
an integrable quaternionic structure. This leads to a formalism 
that parallels very closely that of 2--dimensional conformal field theory 
on Riemann surfaces. In this framework, the notion of  
Fueter analyticity, the quaternionic analogue of complex analyticity, 
plays an essential role. Conformal fields appear as sections of appropriate 
either harmonic real or Fueter holomorphic quaternionic 
line bundles. In the free case, the field equations are statements of either 
harmonicity or Fueter holomorphicity of the relevant conformal fields. 
We obtain compact quaternionic expressions of such basic objects 
as the energy--momentum tensor and the gauge currents for some basic models
in terms of Kulkarni geometry.
We also find a concise expression of the conformal anomaly and a quaternionic 
4--dimensional analogue of the Schwarzian derivative describing the 
covariance of the quantum energy--momentum tensor. Finally, we analyse 
the operator product expansions of free fields.   
\par\noindent
1991 MSC no.: 81T13, 81T90, 81T30
\vfill\eject
\item{\bf 0.} {\bf Introduction}
\vskip.4cm
\par
The success of 2--dimensional conformal field theory both in the study 
of critical 2--dimensional statistical mechanics and perturbative 
string theory is well known \ref{1--3}. Higher dimensional conformal field 
theory is similarly relevant in critical higher dimensional sta\-ti\-sti\-cal 
physics and may eventually play an important role in membrane theory
\ref{4-6}. Unfortunately, so far it has failed to be as fruitful as its 
2--dimensional counterpart in spite of its considerable physical interest.
 
The basic reason of this failure is well--known. In 2 dimensions,
the conformal algebra is infinite dimensional and thus it strongly 
constraints the underlying field theory. It is precisely this that renders 
2--dimensional conformal field theory very predictive and computationally 
efficient. In $d>2$ dimensions, the conformal algebra is instead only 
$(d+1)(d+2)/2$ dimensional and has therefore limited structural implications.
There are however other features of 2--dimensional conformal field theory, 
which turn out to be of considerable salience and may generalize to higher 
dimensions.

In a 2--dimensional conformal model
on an oriented Riemann surface $\Sigma$, the scale of the 
background metric in the action can be absorbed into a multiplicative 
redefinition of the dynamical fields by an appropriate power 
of the scale. The action can then be expressed entirely in terms of the
underlying conformal geometry. The fields become either 
functions or sections of certain holomorphic line bundles on $\Sigma$. 
In the free case, the field equations reduce to the condition of either 
harmonicity or holomorphicity of the fields. 
Complex analyticity is therefore a distinguished feature of these 
field theoretic models allowing the utilization of powerful methods of 
complex analysis such as the Cauchy integral formula and the Laurent 
expansion theorem.

In a higher dimensional conformal model on a manifold $X$, the scale 
of the background metric in the action can be similarly absorbed into a 
multiplicative redefinition of the fields by some power 
of the scale and the action is again expressible entirely in terms of the
underlying conformal geometry, as in the 2--dimensional case.
One may wonder if there are higher dimensional generalizations of 
2--dimensional complex analyticity of the same salience.  
The present paper aims to show that this is in fact so in 4 dimensions. 
The form of analyticity relevant to the 4--dimensional case 
is Fueter's quaternionic analyticity.
This is stronger than real analyticity, as complex analyticity is, and 
yet is weak enough to be fulfilled by a wide class of functions.  
It also allows for a straightforward generalization of 
the main fundamental theorems of complex analysis \ref{7}.

By definition, a complex function $f(z)$ of a complex variable 
$z$ is holomorphic if it satisfies the well-known
Cauchy--Riemann equations $\partial_{\bar z}f=0$. 
Similarly, a quaternionic function $f(q)$ of a quaternionic variable 
$q$ is right (left) Fueter holomorphic if it satisfies the right (left)
Cauchy--Fueter equation $f\partial_{\bar qR}=0$ ($\partial_{\bar qL}f=0$)
\ref{7}, where 
$$
f\partial_{\bar qR}={1\over 4}(\partial_{x0}f+\partial_{xr}fj_r), 
\quad
\partial_{\bar qL}f={1\over 4}(\partial_{x0}f+j_r\partial_{xr}f).
\eqno(0.1)
$$
for $q=x^0+x^rj_r$ with $x^0$, $x^r$ real, $j_r$, $r=1$, 2, 3
being the standard generators of the quaternion field $\Bbb H$.
Here, due to the non commutative nature of $\Bbb H$,
one distinguishes between left and right Fueter analyticity.

We know that Riemann surfaces are the largest class of 2--folds
allowing for global notions of complex analyticity.
It is therefore natural to look for the largest
class of 4--folds on which Fueter analyticity can be 
similarly globally defined.

The closest 4--dimensional analog of a Riemann surface is 
a Kulkarni 4--fold. A Kulkarni 4--fold $X$ is a real 4--fold 
admitting an atlas of quaternionic coordinates $q$ transforming  
as 
$$
q_\alpha=(a_{\alpha\beta}q_\beta+b_{\alpha\beta})
(c_{\alpha\beta}q_\beta+d_{\alpha\beta})^{-1}
\eqno(0.2)
$$
for some constant matrix $\Big(\matrix{a_{\alpha\beta}&b_{\alpha\beta}\cr 
c_{\alpha\beta}&d_{\alpha\beta}\cr}\Big)
\in\GL(2,\Bbb H)$ \ref{8}. As the 2--dimensional
projective quaternionic group $\PGL(2,\Bbb H)$ is isomorphic to 
the orientation preserving 4--dimensional conformal group $\SO_0(5,1)$,
a Kulkarni 4--fold is just an oriented 
real 4--fold with a conformal structure, 
much in the same way as a Riemann surface is an oriented 2--fold with a 
conformal structure.  Note the analogy of the transformations (0.2) with 
the well--known complex Moebius transformations. In 2 dimensions,
Moebius coordinates are just one of infinitely many choices of coordinates
compatible with the underlying conformal structure. In 4 dimensions, 
the quaternionic coordinates $q$ are conversely the only possible choice
\ref{9}.

The Fueter operators (0.1) appear naturally in the geometry of Kulkarni 
4--folds. One can construct a Fueter complex 
$(\Omega^0(X,\zeta_*),\delta)$, where the $\zeta_p$ are certain 
quaternionic line bundles on $X$ 
and $\delta$ is a differential built out of $\partial_q$ 
and $\partial_{\bar q}$, and show its equivalence to the standard 
de Rham complex $(\Omega^*(X),d)$. Exploiting this property, one can show 
that the spaces of closed (anti)selfdual 2--forms, which are two
fundamental invariants of every real 4--fold with an oriented conformal 
structure,  are defined by a condition of right (left) Fueter 
holomorphicity. Kulkarni 4--folds can be further 
equipped with a harmonic real line bundle $\rho$ and, in the spin case, 
with two right/left Fueter holomorphic quaternionic line bundles 
$\varpi^\pm$. The actions of the flat d'Alembertian 
$\square=\star 1\partial_{\bar q}\partial_q$ 
on real sections of $\rho$ and of the Fueter operator
$\bar \partial_R=\partial_{\bar qR}d\bar q$ 
($\bar \partial_L=d\bar q\partial_{\bar qL}$)
on quaternionic sections of $\varpi^+$ ($\varpi^-$) are 
therefore globally defined. These line bundles and operators are of 
considerable salience because of their relation with the conformal
d'Alembertian and the Dirac operator, respectively. 
All the above indicates that Fueter analyticity is 
a natural notion of regularity on Kulkarni 4--folds. 

The family of Kulkarni 4--fold is very vast. It contains such basic 
examples as $S^4$ and $T^4$ and topologically very complicated 4--folds as 
the oriented 4--dimensional Clifford-Klein forms $\Gamma\backslash\Bbb R^4$,  
and $\Gamma\backslash B_1(\Bbb R^4)$, the oriented 4--dimensional Hopf 
manifolds $\Gamma\backslash(S^1\times S^3)$ and the flat sphere bundles on 
a Riemann surface $B_1(\Bbb R^2)\times_G S^2$. Note that all the above
4--folds, like all oriented Riemann surfaces, are Kleinian manifolds. 

A Kulkarni 4--fold $X$ is naturally endowed with a canonical conformal class 
of locally conformally flat metrics. These are the natural metrics for $X$.
The Riemann 2--form, the Ricci 1-form and the Ricci scalar of such metrics 
and all the objects derived from them have particularly simple compact 
expressions in terms of the scale of the metric and the underlying Kulkarni 
structure. Exploiting Fueter calculus, one can also 
derive the general structure of Einstein locally conformally flat metrics, 
when they exist.
Note once more the analogy with the geometry of Riemann surfaces. 
However, while, in the case of Riemann surfaces, every metric is 
automatically locally conformally flat and Einstein, the same is 
no longer true in the case of Kulkarni 4--folds.

On a Kulkarni 4-fold $X$ equipped with a compatible locally conformally flat 
metric, the analogy of the geometry of 4-- and 2--dimensional conformal
field theory becomes manifest. The fields appear as sections of
either $\rho$ or $\varpi^\pm$ or derived line bundles and 
the action can be expressed fully in the language of Kulkarni geometry.
For instance, the action of the standard conformal complex boson model with 
$s/6$ coupling can be cast as 
$$
I(\phi,\bar\phi_{\rm c})=-{8\over \pi^2}\int_X\bar\phi_{\rm c}\square\phi,
\eqno(0.3)
$$
with $\phi$ a complex section of $\rho$. The field equations of $\phi$ 
read simply as 
$\square\phi=0$ and thus imply the harmonicity of $\phi$.
Similarly, the action of the standard massless Dirac fermion model 
can be cast as
$$
\eqalignno{
I(\psi^+,\psi^-,\tilde\psi^+,\tilde\psi^-)
&={2\over \pi^2}
\real\int_X\Big[\tilde\psi^+\bar\partial_R\wedge\star\hskip 2pt 
dq\hskip 2pt \psi^--\psi^+\bar\partial_R\wedge\star\hskip 2pt 
dq\hskip 2pt \tilde\psi^-\Big]
&(0.4)\cr
&={2\over \pi^2}
\real\int_X\Big[\tilde\psi^+\star dq\wedge\bar\partial_L\psi^-
-\psi^+\star dq\wedge\bar\partial_L\tilde\psi^-\Big].
&\cr}
$$
with $\psi^\pm$ complex Grassman sections of $\varpi^\pm$.
$\star$ is similar to the Hodge star, but it depends only 
on the Kulkarni geometry of $X$.
The field equations of $\psi^+$ ($\psi^-$) read as $\psi^+\bar\partial_R=0$
($\bar\partial_L\psi^-=0$) and imply the right (left)
Fueter holomorphicity of $\psi^+$ ($\psi^-$). 
The energy--momentum tensor and the gauge currents have similarly 
simple expressions and geometrically clear properties in this formalism. 

In the quantum case, the operator product expansions of the
quantum fields may be formulated and analyzed exploiting harmonicity and 
Fueter holomorphicity, in a way very close in spirit to the 
analogous approach of 2--dimensional conformal field theory. 
One can further define a quaternionic conformally invariant 
quantum energy--momentum 
tensor $T_{\rm e}$. The Ward identity obeyed by this can be expressed
in terms of the underlying Kulkarni geometry in the form
$$
d\star T_{\rm e}=0
\eqno(0.5)
$$
up to contact terms. Under a coordinate change of the form (0.2),
$T_{\rm e}$ transforms as 
$$
T_{{\rm e}\alpha}=\zeta_{3\alpha\beta}
\big(T_{{\rm e}\beta}+\varrho_{\alpha\beta}\big). 
\eqno(0.6)
$$
Here,  $\varrho_{\alpha\beta}$ depends only on the 
underlying conformal geometry. So, the matching relation (0.6) is completely
analogous to that of the conformally invariant energy--momentum tensor 
in 2--dimensional conformal field theory and $\varrho_{\alpha\beta}$
is a 4--dimensional generalization of the Schwarzian derivative.

The present paper is an attempt at generalizing some of the powerful 
techniques of 2--dimensional conformal field theory to higher dimensional
field theory in a geometric perspective. It is similar in spirit to
but quite different in approach from the work of refs. \ref{10--11}.

In sections 1, 2 and 3, we provide a detailed account of the quaternionic 
geometry of Kulkarni 4--folds   
in a way that parallels as much as possible the standard treatment of the 
geometry of Riemann surfaces. In sections 4 and 5, we analyze the 
geometric properties of a 4--dimensional conformal field theory 
on a Kulkarni 4--fold respectively in the classical and quantum case.
In section 6, we provide a brief outlook of future developments
\par\vskip.6cm
\item{\bf 1.} {\bf Quaternionic linear algebra and group theory}
\vskip.4cm
\par
In this paper, we argue that the geometry underlying 4--dimensional 
conformal field theory is quaternionic. In this section, we review
briefly basic facts of quaternionic linear algebra and group theory.

The quaternion field $\Bbb H$ is the non commutative field generated over
$\Bbb R$ by the symbols $1$ and $j_r$, $r=1$, 2, 3, subject to the relation 
\footnote{}{}
\footnote{${}^1$}{In this paper, we adopt the following conventions. The 
early Latin indices $a$ through $d$ and middle 
Latin indices $i$ through $m$ take the values 0, 1, 2, 3.
The middle Latin indices $e$ through $g$ and the 
late Latin indices $r$ through $v$ take the values 1, 2, 3.
Sum over repeated indices is understood unless they appear on both sides 
of the same identity.}
$$
j_rj_s=-\delta_{rs}+\epsilon_{rst}j_t.
\eqno(1.1)
$$
Hence, a generic quaternion $a\in\Bbb H$ can be written as
$$
a=a^0+a^rj_r, \quad a^0,~a^r\in \Bbb R.
\eqno(1.2)
$$
Quaternionic conjugation is defined by
$$
\bar a=a^0-a^rj_r.
\eqno(1.3)
$$
The real and imaginary parts of a quaternion $a\in \Bbb H$ are defined,
in analogy to the complex case, as
$$
\real a=(1/2)(a+\bar a)=a^0, \quad \imag a=(1/2)(a-\bar a)=a^rj_r.
\eqno(1.4)
$$
The absolute value of a quaternion $a\in \Bbb H$ is given by
$$
|a|=(\bar aa)^{1\over 2}=a^0a^0+a^ra^r.
\eqno(1.5)
$$

The space $\Bbb H^n$ can be given the structure of right 
$\Bbb H$ linear space in natural fashion. Further, 
it can be equipped with the right sesquilinear scalar product
defined by
$$
\langle u, v\rangle=\sum_{k=1}^n\bar u_k v_k,\quad u,~v\in\Bbb H.
\eqno(1.6)
$$

The $n$--dimensional quaternionic general linear group $\GL(n,\Bbb H)$
is the group of invertible $n$ by $n$ 
matrices with entries in $\Bbb H$. Any $T\in \GL(n,\Bbb H)$ defines 
by left matrix action a right $\Bbb H$ linear operator on $\Bbb H^n$.
The $n$--dimensional symplectic group $\Sp(n)$ is the 
subgroup of $\GL(n,\Bbb H)$ formed by those operators 
leaving the scalar product (1.6) invariant. 

$\Bbb H\Bbb P^n$, the $n$ dimensional quaternionic projective space,
is the quotient of $\Bbb H^{n+1}-\{0\}$ by the right multiplicative 
action of the group $\Bbb H_\times$ of non zero quaternions.

The group $\PGL(n+1,\Bbb H)$ is defined as
$$
\PGL(n+1,\Bbb H)=\GL(n+1,\Bbb H)/\Bbb R_\times,
\eqno(1.7)
$$
where $\Bbb R_\times$ is embedded in $\GL(n+1,\Bbb H)$ as the subgroup
$\Bbb R_\times 1_{n+1}$.
$\PGL(n+1,\Bbb H)$ acts on $\Bbb H\Bbb P^n$ by linear fractional 
transformations.

The case $n=1$ will be of special relevance in what follows.
$\GL(1,\Bbb H)$ is simply the group of non zero
quaternions, i. e. $\GL(1,\Bbb H)\cong\Bbb H_\times$.
$\Sp(1)$ is the group of quaternions of unit absolute
value, so $\Sp(1)\cong \SU(2)$. 

(1.2) defines an isomorphism $\Bbb R^4\cong \Bbb H^1$.
Under such an identification, one has \ref{12}
$$
\Spin(4)\cong\Sp(1)\times\Sp(1),
\eqno(1.8)
$$
$$
\SO(4)\cong(\Sp(1)\times\Sp(1))/\Bbb Z_2,
\eqno(1.9)
$$
where ${\Bbb Z}_2$ is embedded in $\Sp(1)\times\Sp(1)$
as $\{\pm(1_1,1_1)\}$.

Similarly, $S^4\cong\Bbb H\Bbb P^1$. The group of orientation preserving
conformal transformations of $S^4$ is the connected component of the 
identity
of $\SO(5,1)$, $\SO_0(5,1)$. The following fundamental isomorphism
holds \ref{8--9}
$$
\SO_0(5,1)\cong\PGL(2,\Bbb H).
\eqno(1.10)
$$
Explicitly, the action of $\PGL(2,\Bbb H)$ on $\Bbb H\Bbb P^1$ is 
given by 
$$
\eqalignno{
T(a)&=(T_{11}a+T_{12})(T_{21}a+T_{22})^{-1},
\hskip 2.7cm  a\in \Bbb H\Bbb P^1
&(1.11)\cr
&=(-aT^{-1}{}_{21}+T^{-1}{}_{11})^{-1}(aT^{-1}{}_{22}-T^{-1}{}_{12}),
&\cr}
$$
for $T\in\PGL(2,\Bbb H)$. 
The above isomorphism fails to hold in $4n$ dimensions
with $n>1$, since in fact $\SO_0(4n+1,1)\not\cong\PGL(n+1,\Bbb H)$.
This is why the 4--dimensional case is so special.
\par\vskip.6cm
\item{\bf 2.} {\bf The Kulkarni 4--folds}
\vskip.4cm
\par
In this paper, we argue that 4 dimensional conformal field theory is 
formulated most naturally on a class of 4-folds admitting an 
integrable quaternionic structure, the Kulkarni 4--folds. 
In the first part of this section, we discuss the local and global 
quaternionic geometry of such 4--folds. We define the Fueter
complex and show its equivalence to the De Rham complex.
In the second part, we introduce the natural differential operators
of a Kulkarni 4--fold, the d'Alembertian and the Fueter operators, 
and show their global definition. 
In the third and final part, we illustrate several basic examples. 

{\it Local quaternionic differential geometry of real 4--folds}

Let $X$ be a real 4--fold. Let $x$ be a local coordinate of $X$ of domain 
$U$. The four components $x^i$, $i=0$, 1, 2, 3, of $x$ can be assembled 
into a quaternionic coordinate $q$ of the same domain given by
$$
q=x^0+x^rj_r.
\eqno(2.1)
$$

The coordinate vector fields $\partial_{xi}$, $i=0$, 1, 2, 3, 
can be similarly organized into a quaternionic vector field 
$\partial_q$ given by 
$$
\partial_q={1\over 4}(\partial_{x0}-\partial_{xr}j_r).
\eqno(2.2)
$$
Also, $\partial_{\bar q}=\overline{\partial_q}$.
$\partial_q$ is a quaternionic differential operator, called Fueter operator,
acting on the space of smooth $\Bbb H$--valued functions $f$ on $U$. 
Since the quaternion field is not commutative, one must distinguish a left 
and a right action of $\partial_q$:
$f\partial_{qR}=(1/4)(\partial_{x0}f-\partial_{xr}fj_r)$ and  
$\partial_{qL}f=(1/4)(\partial_{x0}f-j_r\partial_{xr}f)$.
If $f$ is $\Bbb R$--valued, then $f\partial_{qR}=\partial_{qL}f\equiv
\partial_{q}f$.
$f$ is right (left) Fueter holomorphic if $f\partial_{\bar qR}=0$
($\partial_{\bar qL}f=0$).

The linearly independent wedge products 
$dx^{i_1}\wedge\cdots\wedge dx^{i_p}$ with $0\leq i_1<\cdots<i_p\leq 3$
and $1\leq p\leq 4$ can similarly be assembled into a distinguished set of 
alternate wedge products of the differentials $dq$ and $d\bar q$:
$$
dq=dx^0+dx^rj_r, \vphantom{1\over 2}
\eqno(2.3)
$$
$$
-{1\over 2}dq\wedge d\bar q=
\Big(dx^0\wedge dx^t+{1\over 2}\epsilon_{rst}dx^r\wedge dx^s\Big)j_t,
\eqno(2.4)
$$
$$
+{1\over 2}d\bar q\wedge dq=
\Big(dx^0\wedge dx^t-{1\over 2}\epsilon_{rst}dx^r\wedge dx^s\Big)j_t,
$$
$$
{1\over 6}dq\wedge d\bar q\wedge dq=
dx^1\wedge dx^2\wedge dx^3
-{1\over 2}\epsilon_{rst}dx^0\wedge dx^r\wedge dx^sj_t,
\eqno(2.5)
$$
$$
-{1\over 24}dq\wedge d\bar q\wedge dq\wedge d\bar q
=+{1\over 24}d\bar q\wedge dq\wedge d\bar q\wedge dq=
dx^0\wedge dx^1\wedge dx^2\wedge dx^3.
\eqno(2.6)
$$
All the other combinations of $dq$ and $d\bar q$ of the same type 
can be obtained from these by conjugation. Denoting by 
$\star$ the Hodge star operator with respect to the flat
metric $h=dx^i\otimes dx^i$ on $U$, one has
$$
-{1\over 2}dq\wedge d\bar q=\star\Big(-{1\over 2}dq\wedge d\bar q\Big),
\quad
{1\over 2}d\bar q\wedge dq=-\star\Big({1\over 2}d\bar q\wedge dq\Big),
\eqno(2.7)
$$
$$
{1\over 6}dq\wedge d\bar q\wedge dq=\star dq,
\eqno(2.8)
$$
$$
-{1\over 24}dq\wedge d\bar q\wedge dq\wedge d\bar q
=+{1\over 24}d\bar q\wedge dq\wedge d\bar q\wedge dq=\star 1.
\eqno(2.9)
$$

For any $p$--form $\omega={1\over p!}\omega_{i_1\cdots i_p}
dx^{i_1}\wedge\cdots\wedge dx^{i_p}$ on $U$ with $1\leq p\leq 4$, 
one defines the quaternionic components of $\omega$ by:
$$
\eqalignno{
\omega_q&=\omega(\partial_q)=
{1\over 4}\Big(\omega_0-\omega_rj_r\Big),
&p=1,\hskip1.cm(2.10)\cr
\omega_{\bar qq}&=\omega(\partial_{\bar q},\partial_q)
=-{1\over 8}\Big(\omega_{0r}+{1\over 2}\epsilon_{rst}\omega_{st}\Big)j_r,
&p=2,\hskip1.cm(2.11)\cr
\omega_{q\bar q}&=\omega(\partial_q,\partial_{\bar q})
=+{1\over 8}\Big(\omega_{0r}-{1\over 2}\epsilon_{rst}\omega_{st}\Big)j_r,
&\cr
\omega_{q\bar qq}&=\omega(\partial_q,\partial_{\bar q},\partial_q)
=-{3\over 32}\Big(\omega_{123}+{1\over 2}\epsilon_{rst}\omega_{0st}j_r\Big), 
~~~~~&p=3,\hskip1.cm(2.12)\cr
\omega_{\bar qq\bar qq}&=\omega(\partial_{\bar q},\partial_q,
\partial_{\bar q},\partial_q)
=-{3\over 32}\omega_{0123},
&p=4,\hskip1.cm(2.13)\cr
\omega_{q\bar qq\bar q}&=\omega(\partial_q,\partial_{\bar q},
\partial_q,\partial_{\bar q})
=+{3\over 32}\omega_{0123}.
&\cr}
$$
The remaining components are $\omega_{\bar q}=\omega(\partial_{\bar q})$, 
$p=1$, and $\omega_{\bar qq\bar q}=
\omega(\partial_{\bar q},\partial_q,\partial_{\bar q})$, $p=3$, and are
obtained by conjugation: $\omega_{\bar q}=\overline{\omega_q}$
and $\omega_{\bar qq\bar q}=-\overline{\omega_{q\bar qq}}$.
One can express $\omega$ in terms of its components as follows:
$$
\eqalignno{
\omega&=4\real(\omega_qdq),\vphantom{1\over 2}
&p=1,\hskip1.cm(2.14)\cr
\omega&=
-2\real(\omega_{\bar qq}dq\wedge d\bar q)
-2\real(\omega_{q\bar q}d\bar q\wedge dq),\vphantom{1\over 2}
&p=2,\hskip1.cm(2.15)\cr
\omega&=-{16\over 9}\real(\omega_{q\bar qq} dq\wedge d\bar q\wedge dq),  
&p=3,\hskip1.cm(2.16)\cr
\omega&
={4\over 9}\omega_{\bar qq\bar qq} dq\wedge d\bar q\wedge dq\wedge d\bar q
={4\over 9}\omega_{q\bar qq\bar q} d\bar q\wedge dq\wedge d\bar q\wedge dq, 
\hskip 2.cm
&p=4.\hskip1.cm(2.17)\cr}
$$
Note that, when $p=2$, $\omega$ is $\star$--selfdual ($\star$--antiselfdual)
if and only if $\omega_{q\bar q}=0$ ($\omega_{\bar qq}=0$).

From (2.10)--(2.13), for any $p$--form $\omega$ on $U$ 
with $0\leq p\leq3$, one has 
$$
\eqalignno{
(d\omega)_q&=\omega\partial_{qR}=\partial_{qL}\omega,\vphantom{1\over 2}
&p=0,\hskip1.cm(2.18)\cr
(d\omega)_{\bar qq}&=\partial_{\bar qL}\omega_q-\omega_{\bar q}\partial_{qR},
\vphantom{1\over 2}
&p=1,\hskip1.cm(2.19)\cr
(d\omega)_{q\bar q}&=\partial_{qL}\omega_{\bar q}-\omega_q\partial_{\bar qR},
\vphantom{1\over 2}
&\cr
(d\omega)_{q\bar qq}&={3\over 2}\partial_{qL}\omega_{\bar qq}
+{3\over 2}\omega_{q\bar q}\partial_{qR},
&p=2,\hskip1.cm(2.20)\cr
(d\omega)_{\bar qq\bar qq}&=
2(\partial_{\bar qL}\omega_{q\bar qq}-\omega_{\bar qq\bar q}\partial_{qR}),
\vphantom{1\over 2}
&p=3,\hskip1.cm(2.21)\cr
(d\omega)_{q\bar qq\bar q}&=
2(\partial_{qL}\omega_{\bar qq\bar q}-\omega_{q\bar qq}\partial_{\bar qR}).
\vphantom{1\over 2}
&\cr}
$$
These identities show the relation between the de Rham differential $d$ and 
the Fueter operator $\partial_q$.  

{\it Kulkarni 4--folds}

The Kulkarni $4n$--folds are the real $4n$--folds uniformized by 
$(\Bbb H\Bbb P^n,\PGL(n+1,\Bbb H))$ \ref{8}. This condition turns out 
to be very restrictive. We are interested in the case where $n=1$.

A Kulkarni 4--fold $X$ is a real 4--fold 
\footnote{}{}
\footnote{${}^2$}{In this paper, we shall assume, unless otherwise stated,
that a manifold has no boundary.} 
with an atlas
$\{(U_\alpha,q_\alpha)\}$ of quaternionic coordinates such that, for
$U_\alpha\cap U_\beta\not=\emptyset$, there is $T_{\alpha\beta}\in
\PGL(2,\Bbb H)$ such that
$$
q_\alpha=T_{\alpha\beta}(q_\beta), 
\eqno(2.22)
$$
where the right hand side is given by (1.11).

{\it Global quaternionic differential geometry of Kulkarni 4--folds}

Let $X$ be a Kulkarni 4--fold. 
The local quaternionic tensorial structures defined on each patch 
$U_\alpha$ of $X$, as described above, have
very simple covariance properties under the coordinate transformations 
(2.22), as we shall illustrate next.

For $U_\alpha\cap U_\beta\not=\emptyset$, we define
the matching functions
$$
\eta^+{}_{\alpha\beta}
=-q_\alpha T_{\alpha\beta 21}+T_{\alpha\beta 11},
\quad
\eta^-{}_{\alpha\beta}
=T_{\alpha\beta 21}q_\beta+T_{\alpha\beta 22}.
\eqno(2.23)
$$
The $\eta^\pm{}_{\alpha\beta}$ are nowhere vanishing
on $U_\alpha\cap U_\beta$ since, as will be shown in a moment,
the invertible matching operators of the basic quaternionic tensorial 
structures are polynomial in such objects. 

The matching relation of the vector fields $\partial_{q\alpha}$ 
of eq. (2.2) is
$$
\partial_{q\alpha}=
\eta^-{}_{\alpha\beta}\partial_{q\beta}
(\eta^+{}_{\alpha\beta})^{-1}.
\eqno(2.24)
$$

\vskip.12cm\par\noindent
{\it Proof}. By differentiating (2.22) using (1.11), one gets (2.25) below, 
from which one reads off the identity
$\partial_{x\beta i}x_\alpha{}^0+\partial_{x\beta i}x_\alpha{}^sj_s
=\eta^+{}_{\alpha\beta}(\delta_{0i}+\delta_{ri}j_r)
(\eta^-{}_{\alpha\beta})^{-1}$. Using this relation and (2.2), it is 
straightforward to derive (2.24). \hfill $QED$ \vskip.12cm

The matching relations of the wedge products (2.3)--(2.6) are
$$
dq_\alpha=\eta^+{}_{\alpha\beta}dq_\beta(\eta^-{}_{\alpha\beta})^{-1},
\eqno(2.25)
$$
$$
dq_\alpha\wedge d\bar q_\alpha=
|\eta^+{}_{\alpha\beta}|^2|\eta^-{}_{\alpha\beta}|^{-2}
\eta^+{}_{\alpha\beta}dq_\beta\wedge d\bar q_\beta
(\eta^+{}_{\alpha\beta})^{-1},
\eqno(2.26)
$$
$$
d\bar q_\alpha\wedge dq_\alpha=
|\eta^+{}_{\alpha\beta}|^2|\eta^-{}_{\alpha\beta}|^{-2}
\eta^-{}_{\alpha\beta}d\bar q_\beta\wedge dq_\beta
(\eta^-{}_{\alpha\beta})^{-1},
$$
$$
dq_\alpha\wedge d\bar q_\alpha\wedge dq_\alpha=
|\eta^+{}_{\alpha\beta}|^2|\eta^-{}_{\alpha\beta}|^{-2}
\eta^+{}_{\alpha\beta}
dq_\beta\wedge d\bar q_\beta\wedge dq_\beta
(\eta^-{}_{\alpha\beta})^{-1},
\eqno(2.27)
$$
$$
dq_\alpha\wedge d\bar q_\alpha\wedge dq_\alpha\wedge d\bar q_\alpha
=|\eta^+{}_{\alpha\beta}|^4|\eta^-{}_{\alpha\beta}|^{-4}
dq_\beta\wedge d\bar q_\beta\wedge dq_\beta\wedge d\bar q_\beta,
\eqno(2.28)
$$
$$
d\bar q_\alpha\wedge dq_\alpha\wedge d\bar q_\alpha\wedge dq_\alpha
=|\eta^+{}_{\alpha\beta}|^4|\eta^-{}_{\alpha\beta}|^{-4}
d\bar q_\beta\wedge dq_\beta\wedge d\bar q_\beta\wedge dq_\beta.
$$
The Hodge star operators $\star_\alpha$ associated with 
the flat metrics $h_\alpha$ defined above (2.7) match as 
$$
\star_\alpha=(|\eta^+{}_{\alpha\beta}|
|\eta^-{}_{\alpha\beta}|^{-1})^{-2(p-2)}\star_\beta
\quad {\rm on~}p{\rm -forms}.
\eqno(2.29)
$$

\vskip.12cm\par\noindent
{\it Proof}. (2.25) follows immediately from differentiating (2.22)
using (1.11). (2.26)--(2.28) are trivial consequences of (2.25).
(2.29) follows from comparing (2.26)--(2.28) with (2.7)--(2.9).  
\hfill $QED$ \vskip.12cm

The collection $T=\{T_{\alpha\beta}\}$
associated with the coordinate changes (2.22) defines a flat 
$\PGL(2,\Bbb H)$ 1--cocycle on $X$. In general, this cocycle cannot
be lifted to $\GL(2,\Bbb H)$ by choosing suitable $\GL(2,\Bbb H)$  
representatives of the $T_{\alpha\beta}\in\PGL(2,\Bbb H)$.
One has instead a relation of the form
$$
T_{\alpha\gamma}=w_{\alpha\beta\gamma}T_{\alpha\beta}T_{\beta\gamma},
\quad w_{\alpha\beta\gamma}=\pm 1,
\eqno(2.30)
$$
whenever $U_\alpha\cap U_\beta\cap U_\gamma\not=\emptyset$, where
$w=\{w_{\alpha\beta\gamma}\}$ is a flat $\Bbb Z_2$ 2--cocycle on $X$.

\vskip.12cm\par\noindent
{\it Proof}. Since $T$ is a flat 
$\PGL(2,\Bbb H)$ 1--cocycle on $X$ and the center of 
$\PGL(2,\Bbb H)$ is $\Bbb R_\times 1_2$, (2.30) holds with 
$w$ a flat $\Bbb R_\times$ 2-cocycle on $X$,
by a standard theorem of obstruction theory. From here, using 
(2.23), one can show that relation (2.31) below holds.
Now, set $\phi_{\alpha\beta}
=|\eta^+{}_{\alpha\beta}\eta^-{}_{\alpha\beta}|^{1\over 2}$.
Now, using the relation $T_{\alpha\beta}T_{\beta\alpha}=1_2$,
implied by (2.30),
one can show that either $T_{\alpha\beta 21}\not=0$ and 
$T_{\beta\alpha 21}\not=0$ or $T_{\alpha\beta 21}=0$ and 
$T_{\beta\alpha 21}=0$ and, using (2.23), one can further verify that
$\phi_{\alpha\beta}=(|T_{\alpha\beta 21}|
|T_{\beta\alpha 21}|^{-1})^{1\over 2}$ in the former case and 
$\phi_{\alpha\beta}=(|T_{\alpha\beta 11}|
|T_{\beta\alpha 22}|^{-1})^{1\over 2}$ in the latter case.
$\phi_{\alpha\beta}$ is thus a
positive constant and, from its definition, it is clear that 
$\phi_{\alpha\gamma}=|w_{\alpha\beta\gamma}|
\phi_{\alpha\beta}\phi_{\beta\gamma}$ whenever defined. 
Hence, $|w|=\{|w_{\alpha\beta\gamma}|\}$ is a trivial flat 
$\Bbb R_+$ 2--cocycle on $X$. So, choices can be made so that
$w$ is a $\Bbb Z_2$ 2-cocycle on $X$. \hfill $QED$ \vskip.12cm

From (2.23) and (2.30), it follows that 
$$
\eta^\pm{}_{\alpha\gamma}
=w_{\alpha\beta\gamma}\eta^\pm{}_{\alpha\beta}\eta^\pm{}_{\beta\gamma},
\eqno(2.31)
$$
when $U_\alpha\cap U_\beta\cap U_\gamma\not=\emptyset$. So,
$w$ is the obstruction preventing the smooth $\GL(1,\Bbb H)$ 1--cochain
$\eta^\pm=\{\eta^\pm{}_{\alpha\beta}\}$ on $X$ from being a
1--cocycle. 

Note that $|\eta^\pm|=\{|\eta^\pm{}_{\alpha\beta}|\}$ is 
in any case a smooth $\Bbb R_+$ 1--cocycle.

The flat $\GL(2,\Bbb H)$ 1--cochain $T$ is defined up to a 
redefinition of the form $T_{\alpha\beta}\to
c_{\alpha\beta}T_{\alpha\beta}$, where $c=\{c_{\alpha\beta}\}$
is a flat $\Bbb R_\times$ 1--cocycle. Correspondingly, the 
smooth $\GL(1,\Bbb H)$ 1--cochain $\eta^\pm$ gets 
redefined as $\eta^\pm{}_{\alpha\beta}\to
c_{\alpha\beta}\eta^\pm{}_{\alpha\beta}$,
Now, the flat $\Bbb R_\times$ 1--cocycle $c$ 
can be viewed  canonically as a pair
$(n,a)$, where $n$ and $a$ are respectively a 
flat $\Bbb R_+$ 1--cocycle and a flat 
flat $\Bbb Z_2$ 1--cocycle.
The geometric structures, which we shall construct below, 
are independent from $n$ but do depend on $a$ in general.

Define 
$$
\eqalignno{
\zeta_1&=\eta^-{}_L\otimes\eta^+{}_R,
&(2.32)\cr
\zeta_2^\pm&=|\eta^+{}|^{-2}\otimes|\eta^-{}|^2\otimes
\eta^\pm{}_L\otimes\eta^\pm{}_R|_{\imag\Bbb H},
&(2.33)\cr
\zeta_3&=|\eta^+{}|^{-2}\otimes|\eta^-{}|^2\otimes
\eta^-{}_L\otimes\eta^+{}_R,
&(2.34)\cr
\zeta_4&=|\eta^+{}|^{-4}\otimes|\eta^-{}|^4,
&(2.35)\cr}
$$
where for $u,~v\in\Bbb H_\times\cong\GL(1,\Bbb H)$, $u_L\otimes v_R$ is
the $\Bbb R$ linear operator on $\Bbb H$ defined by  $(u_L\otimes v_R)a
=uav^{-1}$ for $a\in\Bbb H$. Then, $\zeta_1$ and 
$\zeta_3$ are smooth $(\GL(1,\Bbb H)\times\GL(1,\Bbb H))/\Bbb R_\times$ 
1-cocycles, where $\Bbb R_\times$
is embedded into $\GL(1,\Bbb H)\times\GL(1,\Bbb H)$ 
as $\Bbb R_\times(1_1,1_1)$;
$\zeta_2^\pm$ is a smooth $\PGL(1,\Bbb H)$ 1--cocycle; $\zeta_4$ is 
a smooth $\Bbb R_+$ 1--cocycle.

\vskip.12cm\par\noindent
{\it Proof}. This follows readily from the definitions and from (2.31).
\hfill $QED$ \vskip.12cm

Let $\omega\in\Omega^p(X)$ be a $p$--form
\footnote{}{}
\footnote{${}^3$}{Let $V$ be a $\Bbb F$ vector space.
When $\xi$ is a smooth $\GL(V)$ 1--cocycle 
on the non empty open subset $O$ of $X$, we denote by $\Omega^p(O,\xi)$ the 
$\Bbb F$ vector space of $p$--form sections of $\xi$ on $O$. In particular, 
$\Omega^p(O)$ is the space of real $p$--forms on $O$.}. 
Using (2.10)--(2.13), we can
associate with $\omega$ the collection of its local components on the  
coordinate patches $U_\alpha$. 
If $\omega\in\Omega^1(X)$,
 $\omega_q=\{\omega_{q\alpha}\}\in\Omega^0(X,\zeta_1)$ 
and the map $\omega\to\omega_q$ is an $\Bbb R$--linear 
isomorphism of $\Omega^1(X)$ onto $\Omega^0(X,\zeta_1)$. 
On account of (2.29), the spaces $\Omega^{2\pm}(X)$ of 
$\star$--(anti)selfdual 2--forms on $X$ are covariantly defined.
If $\omega\in\Omega^{2+}(X)$,
$\omega_{\bar qq}=\{\omega_{\bar qq\alpha}\}\in
\Omega^0(X,\zeta_2^+)$ and the map $\omega\to\omega_{\bar qq}$ is an 
$\Bbb R$--linear isomorphism of $\Omega^{2+}(X)$ onto $\Omega^0(X,\zeta_2^+)$ 
and, similarly, if $\omega\in\Omega^{2-}(X)$,
$\omega_{q\bar q}=\{\omega_{q\bar q\alpha}\}\in
\Omega^0(X,\zeta_2^-)$ and the map $\omega\to\omega_{q\bar q}$ 
is an $\Bbb R$--linear isomorphism of $\Omega^{2-}(X)$ onto 
$\Omega^0(X,\zeta_2^-)$. 
If $\omega\in\Omega^3(X)$,
$\omega_{q\bar qq}
=\{\omega_{q\bar qq\alpha}\}\in\Omega^0(X,\zeta_3)$ and the map 
$\omega\to\omega_{q\bar qq}$ is an $\Bbb R$--linear isomorphism of 
$\Omega^3(X)$ onto $\Omega^0(X,\zeta_3)$. 
Finally, if $\omega\in\Omega^4(X)$,
$\omega_{\bar qq\bar qq}=\{\omega_{\bar qq\bar qq\alpha}\}\in
\Omega^0(X,\zeta_4)$ and $\omega_{q\bar qq\bar q}
=\{\omega_{q\bar qq\bar q\alpha}\}\in\Omega^0(X,\zeta_4)$ and the maps
$\omega\to\omega_{\bar qq\bar qq}$ and $\omega\to\omega_{q\bar qq\bar q}$ 
are both $\Bbb R$--linear isomorphisms of 
$\Omega^4(X)$ onto $\Omega^0(X,\zeta_4)$. 

\vskip.12cm\par\noindent
{\it Proof}. This follows easily from the definition of the quaternionic 
components of the form $\omega$, given in (2.10)--(2.13), and from (2.24)
upon taking (2.32)--(2.35) into account. For $p=1$, one has 
$\omega_{q\alpha}=\omega(\partial_{q\alpha})
=\omega(\eta^-{}_{\alpha\beta}\partial_{q\beta}
(\eta^+{}_{\alpha\beta})^{-1})=
\eta^-{}_{\alpha\beta}\omega(\partial_{q\beta})
(\eta^+{}_{\alpha\beta})^{-1}=\zeta_{1\alpha\beta}\omega_{q\beta}$.
The proof for the other $p$ values is analogous.
\hfill $QED$ \vskip.12cm

By the above isomorphisms, the standard de Rham complex 
$$
\matrix{   
             &   &   &d^+     &\Omega^{2+}(X)&d       &   &   &   \cr
             &d  &   &\nearrow&              &\searrow&   &d  &   \cr
\Omega^0(X)&\longrightarrow&\Omega^1(X)&   &   &   &
\Omega^3(X)&\longrightarrow&\Omega^4(X)\cr
             &   &   &\searrow&              &\nearrow&   &   &   \cr
             &   &   &d^-     &\Omega^{2-}(X)&d     &     &   &   \cr}
\eqno(2.36)
$$
is equivalent to the Fueter complex

$$
\matrix{   
  &        &   &\delta^+&\Omega^0(X,\zeta_2^+)&\delta  &   &        &   \cr
  &\delta  &   &\nearrow&                     &\searrow&   &\delta  &   \cr
\Omega^0(X)&\longrightarrow&\Omega^0(X,\zeta_1)&   &   &   &
\Omega^0(X,\zeta_3)&\longrightarrow&\Omega^0(X,\zeta_4)\cr
  &        &   &\searrow&                     &\nearrow&   &        &   \cr
  &        &   &\delta^-&\Omega^0(X,\zeta_2^-)&\delta  &   &        &   \cr},
\eqno(2.37)
$$
where the Fueter operators $\delta$ are defined by the right hand sides of 
(2.18)--(2.21), with $\delta^+$ and $\delta^-$ corresponding respectively 
to the first and second expression (2.19). The two definitions of the last
$\delta$ differ only by their sign.

From here, one sees that a 2--form $\omega\in\Omega^{2+}(X)$
($\omega\in\Omega^{2-}(X)$) is closed if and only if 
$\omega_{\bar qq}\partial_{\bar qR}=0$ 
($\partial_{\bar qL}\omega_{q\bar q}=0$), 
that is if and only if $\omega_{\bar qq}$ ($\omega_{q\bar q}$) is right 
(left) Fueter holomorphic. 

\vskip.12cm\par\noindent
{\it Proof}. Let $\omega\in\Omega^{2+}(X)$. Then, $\omega_{q\bar q}=0$.
So, if further $d\omega=0$, one has $\omega_{\bar qq}\partial_{\bar qR}
=-\overline{\partial_{qL}\omega_{\bar qq}}
=-{2\over 3}\overline{(d\omega)_{q\bar qq}}=0$, by (2.11) and (2.20). 
The corresponding statement for a closed $\omega\in\Omega^{2-}(X)$
can be proven in analogous manner. \hfill $QED$ \vskip.12cm

\par\noindent
The spaces of closed (anti)selfdual 2--forms are important invariants of
any real 4--fold endowed with a conformal structure. 
The above proposition shows that, on a Kulkarni 4--fold,
such spaces are defined by a condition of Fueter holomorphicity.
We believe that this result highlights quite clearly the relevance of Fueter 
analyticity to the geometry of Kulkarni 4--folds.

{\it The 1--cocycle} $\rho$ {\it and the d'Alembert operator} $\square$
\footnote{}{}
\footnote{${}^4$}{In the mathematical literature, this operator is usually 
called Laplacian and is denoted by $\Delta$. In the spirit of field theory, 
we rather think of it as the Euclidean version of the d'Alembert
operator $\square$.}

Let $X$ be a Kulkarni 4-fold. We set
$$
\rho=|\eta^+{}|^{-1}\otimes|\eta^-{}|.
\eqno(2.38)
$$
Then, $\rho$ is a smooth $\Bbb R_+$ 1--cocycle.

\vskip.12cm\par\noindent
{\it Proof}. This follows immediately 
from the definition and from (2.31). \hfill $QED$ \vskip.12cm

Let $F\in\Omega^0(X,\rho)$. Set
$$
\square F=\partial_{\bar q}\partial_qF \star 1
\eqno(2.39)
$$
on each coordinate patch. Then, 
$\square F=\{(\square F)_\alpha\}\in\Omega^4(X,\rho^{-1})$. 

\vskip.12cm\par\noindent
{\it Proof}. Let $(U,q)$ be a quaternionic chart of $X$ and let 
$f\in\Omega^0(U)$. Then,
$$
\partial_{\bar q}\partial_q f\star 1={1\over 16}d\star df.
\eqno(2.40)
$$
This relation can be easily checked by 
evaluating the right hand side in terms of the components of the  
real coordinate $x$ contained in $q$ (cf. eq. (2.1)). 
Using (2.29), (2.40) and (2.38) and the matching relation 
$F_\alpha=\rho_{\alpha\beta}F_\beta$, one finds
$$
\partial_{\bar q\alpha}\partial_{q \alpha}F_\alpha \star_\alpha 1
=-\partial_{\bar q\beta}\partial_{q \beta}
(\rho_{\alpha\beta}{}^{-1})F_\beta\star_\beta 1
+\rho_{\alpha\beta}{}^{-1}
\partial_{\bar q\beta}\partial_{q \beta}F_\beta\star_\beta 1.
\eqno(2.41)
$$
Now, using the relation $T_{\alpha\beta}T_{\beta\alpha}=1_2$ implied by 
(2.30), one can show that either $T_{\alpha\beta 21}\not=0$ and 
$T_{\beta\alpha 21}\not=0$ or $T_{\alpha\beta 21}=0$ and 
$T_{\beta\alpha 21}=0$ and, using further (2.23), one can verify that
$\rho_{\alpha\beta}=|T_{\alpha\beta 21}||T_{\beta\alpha 21}|
|q_\beta+T_{\alpha\beta 21}{}^{-1}T_{\alpha\beta 22}|^2$ in the former case 
and $\rho_{\alpha\beta}=|T_{\alpha\beta 22}||T_{\beta\alpha 11}|$ in the 
latter case. Using these expressions, one finds that 
$$
\partial_{\bar q\beta}\partial_{q \beta}
(\rho_{\alpha\beta}{}^{-1})=0
\eqno(2.42)
$$
by direct computation. The statement follows now readily from (2.41) 
and (2.42). \hfill $QED$ \vskip.12cm

Note that, in terms of the real coordinate $x$ contained in $q$, 
$\square F={1\over 16}(\partial_{x0}\partial_{x0}
+\partial_{xr}\partial_{xr})F
\star 1$. So, $\square$ is essentially the euclidean d'Alembertian
operator. (2.42) shows then that the 1-cocycle $\rho$ is harmonic.
This allows for a global definition of harmonicity on 
a Kulkarni 4--fold $X$. An element $F\in\Omega^0(X,\rho)$ is said
harmonic if $\square F=0$. In such a case, $F$ is given locally 
by the real part of some Fueter holomorphic function $K$ \ref{7}.

{\it The 1--cocycles} $\varpi^\pm$ {\it and the operators}
$\bar\partial_{R,L}$

Let $X$ be a Kulkarni 4--fold such that $w=1$. We set
$$
\varpi^+=|\eta^+|^{1\over 2}\otimes
|\eta^-|^{-{3\over 2}}\otimes\eta^+{}_R,
\quad
\varpi^-=|\eta^+|^{-{3\over 2}}\otimes
|\eta^-|^{1\over 2}\otimes\eta^-{}_L,
\eqno(2.43)
$$
where for $u\in\Bbb H_\times\cong\GL(1,\Bbb H)$, $u_R$ ($u_L$) is the  
the left (right) $\Bbb H$ linear operator on $\Bbb H$ defined by  
$u_Ra=au^{-1}$ ($u_La=ua$) for $a\in\Bbb H$.
Then, $\varpi^\pm$ is a smooth $\GL(1,\Bbb H)$ 1--cocycle
on $X$. 

\vskip.12cm\par\noindent
{\it Proof}. This follows readily from (2.31), taking into account that
$w=1$ in this case, by assumption. \hfill $QED$ \vskip.12cm

\par\noindent 
Note that $\varpi^\pm$ depends on the choice of a $\Bbb Z_2$
1--cocycle $a$, as discussed above (2.32). 
We assume that a choice is made once and for all.

For $\Phi\in\Omega^0(X,\varpi^+)$ and $\Psi\in\Omega^0(X,\varpi^-)$, 
we set
$$
\Phi\bar\partial_R
=\Phi\partial_{\bar qR}d\bar q,
\quad
\bar\partial_L\Psi
=d\bar q\partial_{\bar qL}\Psi
\eqno(2.44)
$$
on each coordinate patch.
Then, $\Phi\bar\partial_R=\{(\Phi\bar\partial_R)_\alpha\}
\in\Omega^1(X,\varpi^+)$ and $\bar\partial_L\Psi
=\{(\bar\partial_L\Psi)_\alpha\}\in\Omega^1(X,\varpi^-)$.

\vskip.12cm\par\noindent
{\it Proof}. We show only that
$\Phi\bar\partial_R\in\Omega^1(X,\varpi^+)$, since the proof of 
the corresponding statement for $\Psi$ is totally analogous.
For the rest of the proof, introducing a slightly inconsistent notation, 
we denote by $\varpi^\pm$ the matching functions 
defined by (2.43) with the indices $R$, $L$ suppressed.
Let $(U,q)$ be a quaternionic chart of $X$ and let
$f\in\Omega^0(U)\otimes\Bbb H$. Then, one has
$$
f\partial_{\bar qR}\star 1={1\over 4}df\wedge\star dq.
\eqno(2.45)
$$
(2.45) can be easily be checked by expressing both sides in terms of the
components of the real coordinate $x$ contained in $q$.  
Using (2.45) and the matching relation 
$\Phi_\alpha=\Phi_\beta\varpi^+{}_{\beta\alpha}$, one has
$$
\Phi_\alpha\partial_{\bar q\alpha R}\star_\alpha 1
={1\over 4}d\Phi_\beta\wedge\varpi^+{}_{\beta\alpha}\star_\alpha dq_\alpha
+\Phi_\beta\varpi^+{}_{\beta\alpha}\partial_{\bar q\alpha R}\star_\alpha 1.
\eqno(2.46)
$$
From (2.43) and (2.23), one computes 
$$
\eqalignno{
\varpi^+{}_{\beta\alpha}\partial_{\bar q\alpha R}
&=|\eta^+{}_{\alpha\beta}|^{-{5\over 2}}|\eta^-{}_{\beta\alpha}|^{-{3\over 2}}
\Big\{\overline{\eta^+{}_{\alpha\beta}}\Big[
-{5\over 4}|\eta^+{}_{\alpha\beta}|^{-2}(|\eta^+{}_{\alpha\beta}|^2)
\partial_{\bar q\alpha R}
&(2.47)\cr
&\hphantom{=}
-{3\over 4}|\eta^-{}_{\beta\alpha}|^{-2}(|\eta^-{}_{\beta\alpha}|^2)
\partial_{\bar q\alpha R}\Big]
+\overline{\eta^+{}_{\alpha\beta}}\partial_{\bar q\alpha R}\Big\}
&\cr
&=-{3\over 8}
|\eta^+{}_{\alpha\beta}|^{-{5\over 2}}|\eta^-{}_{\beta\alpha}|^{-{3\over 2}}
\overline{(\eta^-{}_{\beta\alpha})^{-1}\big[
T_{\beta\alpha 21}\eta^+{}_{\alpha\beta}+
\eta^-{}_{\beta\alpha}T_{\alpha\beta 21}\big]}.
&\cr}
$$
Using the relation $T_{\alpha\beta}T_{\beta\alpha}=1_2$, following from 
(2.30), and (2.23), one finds that
$$
T_{\beta\alpha 21}\eta^+{}_{\alpha\beta}+
\eta^-{}_{\beta\alpha}T_{\alpha\beta 21}=0.
\eqno(2.48)
$$
Combining (2.47) and (2.48), one concludes that 
$$
\varpi^+{}_{\beta\alpha}\partial_{\bar q\alpha R}=0.
\eqno(2.49)
$$
From (2.8), (2.27) and (2.43), one verifies further that
$$
{1\over 4}\varpi^+{}_{\beta\alpha}\star_\alpha dq_\alpha
={1\over 4}\star_\beta dq_\beta\varpi^-{}_{\beta\alpha}.
\eqno(2.50)
$$
By (2.46), (2.49) and (2.50), one has, using (2.45), 
$$
\eqalignno{
\Phi_\alpha\partial_{\bar q\alpha R}\star_\alpha 1
&={1\over 4}d\Phi_\beta\wedge\star_\beta dq_\beta\varpi^-{}_{\beta\alpha}
&(2.51)\cr
&=\Phi_\beta\partial_{\bar q\beta R}\varpi^-{}_{\beta\alpha}\star_\beta 1.
&\cr}
$$
From this relation, using (2.43), (2.25) and (2.29) with $p=0$, it is a 
simple matter to check that $(\Phi\bar\partial_R)_\alpha
=(\Phi\bar\partial_R)_\beta\varpi^+{}_{\beta\alpha}$, showing the statement.
\hfill $QED$ \vskip.12cm

(2.49) and its left analog show that the 1--cocycle
$\varpi^+$ ($\varpi^-$) is right (left) Fueter holomorphic.
This allows for a global definition of Fueter holomorphicity
on a Kulkarni 4--fold $X$. An element $\Phi\in\Omega^0(X,\varpi^+)$ 
($\Psi\in\Omega^0(X,\varpi^-)$) is right (left) Fueter holomorphic
if $\Phi\bar\partial_R=0$ ($\bar\partial_L\Psi=0$).

{\it Topological properties of Kulkarni 4--folds}

On account of the isomorphism (1.10), (2.22) entails that 
a Kulkarni 4--fold is just a real 4--fold with an integrable 
oriented conformal structure. 

A Kulkarni 4--fold structure entails a reduction of the structure group of 
$X$ from $\GL(4,\Bbb R)$ to 
$(\GL(1,\Bbb H)\times\GL(1,\Bbb H))/\Bbb R_\times$.

\vskip.12cm\par\noindent
{\it Proof}. Indeed, from (2.24), it appears that the smooth 1--cocycle 
implementing the matching relations in $TX$ is the
$(\GL(1,\Bbb H)\times\GL(1,\Bbb H))/\Bbb R_\times$ 
1-cocycle $\eta^-{}_L\otimes\eta^+{}_R$. \hfill $QED$ \vskip.12cm

\par\noindent
The resulting $(\GL(1,\Bbb H)\times\GL(1,\Bbb H))/\Bbb R_\times$ structure
on $X$, being yielded by coordinates, is integrable.

Since $(\GL(1,\Bbb H)\times\GL(1,\Bbb H))/\Bbb R_\times$ is a connected 
group, $X$ is oriented. Hence, the first Stieffel--Whitney class of $X$ 
vanishes:
$$
w_1(X)=1.
\eqno(2.52)
$$

The flat $\Bbb Z_2$ 2--cocycle $w$ appearing in (2.30) defines 
a cohomology class $w\in H^2(X,\Bbb Z_2)$. It can be seen that $w$ is 
precisely the second Stieffel--Whitney class of $X$:
$$
w_2(X)=w.
\eqno(2.53)
$$

\vskip.12cm\par\noindent
{\it Proof}. $|\eta^+|\otimes |\eta^-|^{-1}$ is a smooth $\Bbb R_+$
1--cocycle, hence, it is trivial. So, the smooth 
$(\GL(1,\Bbb H)\times\GL(1,\Bbb H))/\Bbb R_\times$ 1--cocycle 
$\eta^-{}_L\otimes\eta^+{}_R$ is equivalent to the 
smooth $(\Sp(1)\times\Sp(1))/\Bbb Z_2$ 1--cocycle 
$\theta^-{}_L\otimes\theta^+{}_R$, where $\Bbb Z_2$ is embedded in 
$\Sp(1)\times\Sp(1)$ as $\{\pm(1_1,1_1)\}$ and 
$\theta^\pm=|\eta^\pm|^{-1}\otimes\eta^\pm$ is an $\Sp(1)$
1--cochain. This yields a reduction of the structure group of
$X$ from $(\GL(1,\Bbb H)\times\GL(1,\Bbb H))/\Bbb R_\times$
to $(\Sp(1)\times\Sp(1))/\Bbb Z_2$. Now $\theta^\pm$ satisfies relation 
(2.31) with $\eta^\pm$ substituted by $\theta^\pm$.
From the isomorphisms (1.8) and (1.9), it follows then that 
the $\Bbb Z_2$ 2--cocycle $w$ is precisely the obstruction to 
lifting the structure group of $X$ from $\SO(4)$ to $\Spin(4)$. 
This identifies $w$ as a representative of the 
second Stieffel--Whitney class of $X$.
\hfill $QED$ \vskip.12cm 

\par\noindent 
So, the spin Kulkarni 4--folds are precisely those for which 
$w=1$. In such a case, the spin structures correspond 
precisely to the choices of the $\Bbb Z_2$ 1--cocycle $a$ on $X$
discussed above (2.32). Indeed, as is well-known, such choices 
describe the cohomology group $H^1(X,\Bbb Z_2)$.

As $X$ is endowed with an integrable oriented conformal structure, 
the first Pontryagin class of $X$ is zero:
$$
p_1(X)=0.
\eqno(2.54)
$$

\vskip.12cm\par\noindent
{\it Proof}. The integrability of the conformal structure 
implies the existence of locally conformally flat metrics, whose Weyl
2--form vanishes \ref{9}. The Pontryagin density, which is quadratic in the 
components of $W(g)$, consequently vanishes too. 
\hfill $QED$ \vskip.12cm 

Let $X$ be compact.
As $p_1(X)=0$, the signature of $X$ vanishes as well, $\sigma(X)=0$.
This entails that the Euler characteristic of $X$ is even:
$$
\chi(X)\in 2\Bbb Z.
\eqno(2.55)
$$

If $X$ is compact, then (2.54) and (2.55) imply that $X$ bounds an oriented
5--fold by the Thom Pontryagin theorem \ref{8}. 

All the 1--cocycles defined in the previous subsections yield 
smooth vector bundles on $X$ in the usual manner. In particular, $\zeta_1$ 
and $\zeta_3$ are smooth $(\GL(1,\Bbb H)\times\GL(1,\Bbb H))/\Bbb R_\times$ 
line bundles, the $\zeta_2^\pm$ are smooth $\PGL(1,\Bbb H)$ line bundles, 
$\zeta_4$ and $\rho$ are smooth $\Bbb R_+$ line bundles 
and the $\varpi^\pm$ are $\GL(1,\Bbb H)$ line bundles.

The operator $\square$ is elliptic. Therefore, when $X$ is compact, the 
subspace of the harmonic $F\in\Omega^0(X,\rho)$ is finite dimensional.
The operators $\bar\partial_{R,L}$ are also elliptic. Hence,
if $X$ is compact, the subspace of right (left) Fueter holomorphic
$\Phi\in\Omega^0(X,\varpi^+)$ ($\Psi\in\Omega^0(X,\varpi^-)$)
is similarly finite dimensional. 
In the next section, we shall show that $\square$ 
and the $\bar\partial_{R,L}$ are related respectively to the 
conformal d'Alembertian and to a certain Dirac operator. This 
will allow us to derive vanishing theorems.

{\it Kulkarni automorphisms}

An orientation preserving diffeomorphism $f$ of $X$ is a Kulkarni
automorphism of $X$ if $q_\alpha\circ f\circ q_\beta{}^{-1}$, whenever 
defined, is a restriction of some element of $\PGL(2,\Bbb H)$. The 
Kulkarni automorphisms of $X$ form a group under composition, $\Aut(X)$.

{\it Examples of Kulkarni 4--folds}

The basic example of Kulkarni 4--fold is $\Bbb H\Bbb P^1$. As a 
4--fold $\Bbb H\Bbb P^1\cong S^4$. Indeed, $\Bbb H\Bbb P^1$ can be covered 
by two quaternionic charts $(q_\alpha,U_\alpha)$, $\alpha=1$, 2, where
$U_\alpha=\{(p_1,p_2)\in\Bbb H^2-\{(0,0)\}|p_\alpha\not=0\}/\Bbb H_\times$ 
and $q_1=p_2p_1{}^{-1}$ and $q_2=-p_1p_2{}^{-1}$. One has $q_2=-(q_1)^{-1}$ 
on the overlap $U_1\cap U_2$. Under the isomorphism $\Bbb H^1\cong \Bbb R^4$,
this matching relation is equivalent to that of the customary stereographic 
projection of $S^4$. Clearly, $\Aut(\Bbb H\Bbb P^1)=\PGL(2,\Bbb H)$. 
Also, $w(\Bbb H\Bbb P^1)=1$.

Let $D$ be a simply connected non empty open subset of $\Bbb H\Bbb P^1$.
Then, $D$ is a Kulkarni 4--fold with the Kulkarni structure induced by that
of $\Bbb H\Bbb P^1$. When $D$ is a proper subset of $\Bbb H\Bbb P^1$, 
then $D$ can be covered by a single quaternionic chart $(q,U)$ with $U=D$. 
The automorphism group $\Aut(D)$ of $D$ is the subgroup of $\PGL(2,\Bbb H)$ 
mapping $D$ onto itself. Clearly, $w(D)=1$.

A Kleinian group $\Gamma$ for $D$ is a subgroup of 
$\Aut(D)$ acting freely and properly discontinuously on $D$ \ref{9,13}. 
The Kleinian manifold $\Gamma\backslash D$ is then  
a Kulkarni 4--fold, as it is a real 4--fold uniformized
by $(\Bbb H\Bbb P^1,\PGL(2,\Bbb H))$. $\Aut(\Gamma\backslash D)$ can be 
identified with the normalizer of $\Gamma$ in $\Aut(D)$. 
$w(\Gamma\backslash D)=1$ if and only if 
$\Gamma$ can be lifted to a subgroup of $\GL(2,\Bbb H)$.

We consider next several standard examples.

$i)$ $D=\Bbb H\Bbb P^1$.  $\Aut(\Bbb H\Bbb P^1)=\PGL(2,\Bbb H)$, 
as shown earlier. By a simple argument based on Lefschetz's fixed point 
theorem, it is easy to see that there is no non trivial Kleinian 
group $\Gamma$ for $\Bbb H\Bbb P^1$, since every $T\in\PGL(2,\Bbb H)$ has 
at least a fixed point in $\Bbb H\Bbb P^1$. Thus, there are no
Kulkarni 4--folds covered by $\Bbb H\Bbb P^1$ except for $\Bbb H\Bbb P^1$ 
itself. 

$ii)$ $D=\Bbb H^1$. It appears that $\Bbb H^1\cong\Bbb R^4$, as a 4--fold.  
$\Aut(\Bbb H^1)$ is the subgroup of $\PGL(2,\Bbb H)$ formed by those  
$T$ such that $T_{21}=0$. There are plenty of Kleinian groups $\Gamma$ 
for $\Bbb H^1$. Among these, the orientation preserving 4--dimensional 
Bieberbach groups, which have been classified \ref{13}. In this way, the 
Kulkarni 4--folds $\Gamma\backslash\Bbb H$ covered by $\Bbb H^1$ include 
the 4--torus $T^4$ and the oriented 4--folds finitely covered by it. 

$iii)$ $D=B_1(\Bbb H^1)$. As a 4--fold, $B_1(\Bbb H^1)\cong B_1(\Bbb R^4)$, 
the unit ball in $\Bbb R^4$.  
$\Aut(B_1(\Bbb H^1))$ is the subgroup of $\PGL(2,\Bbb H)$ formed by those 
$T$ such that $|T_{11}|^2-|T_{21}|^2=|T_{22}|^2-|T_{12}|^2=k$ for some
$k\in\Bbb R_+$ and $\bar T_{11}T_{12}-\bar T_{21}T_{22}=0$. There are 
plenty of Kleinian groups $\Gamma$ for $B_1(\Bbb H^1)$. The 
Kulkarni 4--folds $\Gamma\backslash B_1(\Bbb H^1)$ covered by 
$B_1(\Bbb H^1)$ are the 4--dimensional analogue of higher genus Riemann 
surfaces.

$iv)$ $D=\Bbb H^1-\{0\}$. As a 4--fold, $\Bbb H^1-\{0\}\cong 
\Bbb R^4-\{0\}$. $\Aut(\Bbb H^1-\{0\})$ contains as a 
subgroup of index 2 the subgroup of $\PGL(2,\Bbb H)$ formed by those  
$T$ such that $T_{12}=T_{21}=0$. There are plenty of Kleinian groups 
$\Gamma$ for $\Bbb H^1-\{0\}$. Among the Kulkarni 4--folds
$\Gamma\backslash(\Bbb H^1-\{0\})$ covered by $\Bbb H^1-\{0\}$,
there are the oriented 4--dimensional Hopf manifolds, that is 
$S^3\times S^1$ and the oriented compact 4--folds finitely covered by it.

$v)$ $D=\Bbb H^1-\Bbb R^1$. As a 4--fold, $\Bbb H^1-\Bbb R^1
\cong \Bbb R^4-\Bbb R^1$. There are many Kleinian groups $\Gamma$ for
$\Bbb H^1-\Bbb R^1$. The Kulkarni 4--folds $\Gamma\backslash
(\Bbb H^1-\Bbb R^1)$ covered by $\Bbb H^1-\Bbb R^1$ include the 
flat $S^2$ fiber bundle on a compact Riemann surface,
as $\Bbb R^4-\Bbb R^1\cong B_1(\Bbb R^2)\times S^2$.
\par\vskip.6cm
\item{\bf 3.} {\bf The geometry of Kulkarni 4--folds from a Riemannian point 
of view}
\vskip.4cm
\par
4--dimensional conformal field theory is most naturally formulated
in a locally conformally flat metric background. One expects calculations 
to simplify considerably if this background has special properties,
such as having a large group of isometries or being Einstein. 
A Kulkarni 4--fold is equipped with a canonical conformal equivalence class 
of locally conformally flat metrics.
These are studied in the first part of this section using the quaternionic 
geometric framework introduced above. We also derive
conditions for the existence of an Einstein representative in the class
and its general form, when it exists.
In the second part of the section, we show that the operators 
$\square$ and the $\bar\partial_{R,L}$ are related respectively to the 
conformal d'Alembertian and to a certain Dirac operator. This 
will allow us to derive vanishing theorems \`a la Bochner for their 
kernels. Examples are provided in the third and final part 
of the section. 

{\it Local quaternionic Riemannian geometry of a real 4--fold}

Let $X$ be a real 4-fold. Let $x$ be a local coordinate of $X$ of domain
$U$. On $U$, one can define the conformally flat vierbein
$$
e_a=\ee^{-\varphi}\delta_a{}^i\partial_{xi}.
\eqno(3.1)
$$
Its dual vierbein is
$$
e^\vee{}_a=\ee^\varphi\delta_{ai}dx^i.
\eqno(3.2)
$$
The associated metric is
$$
g=e^\vee{}_a\otimes e^\vee{}_a=\ee^{2\varphi}dx^i\otimes dx^i.
\eqno(3.3)
$$

The components of the vierbein $e_a$, $a=0$, 1, 2, 3,
can be assembled into the quaternionic einbein
$$
e=(1/4)(e_0-e_fj_f).
\eqno(3.4)
$$
By (2.2) and (3.1), $e$ is given by
$$
e=\ee^{-\varphi}\partial_q.
\eqno(3.5)
$$
Similarly, the components of the dual vierbein $e^\vee{}_a$, $a=0$, 1, 2, 3,
can be assembled into the quaternionic dual einbein
$$
e^\vee=e^\vee{}_0+e^\vee{}_fj_f.
\eqno(3.6)
$$
From (2.3) and (3.2), one has 
$$
e^\vee=\ee^\varphi dq.
\eqno(3.7)
$$
The metric $g$ is then given by
$$
g=\real(\bar e^\vee\otimes e^\vee)=\ee^{2\varphi}\real(d\bar q\otimes dq).
\eqno(3.8)
$$
The Hodge star operator $*$ of $g$ is related to $\star$ as 
$$
*=\ee^{-2(p-2)\varphi}\star
\quad {\rm on~}p{\rm -forms}.
\eqno(3.9)
$$

Many formulae of Riemannian geometry take a particularly compact form when 
expressed in terms of $e$ and $e^\vee$. Below, we shall adopt the Cartan
formulation of Riemannian geometry.

The components of the spin connection $\omega_{ab}$ 1--form
can be organized into the two quaternionic 1--forms 
$$
\eqalignno{
\omega^+&=-{1\over 4}\Big(\omega_{00}+\omega_{0f}\bar\jmath_f
+\omega_{e0}j_e+\omega_{ef}j_e\bar\jmath_f\Big)
={1\over 2}\Big(\omega_{0g}+{1\over 2}\epsilon_{efg}\omega_{ef}\Big)j_g,
&(3.10)\cr
\omega^-&=+{1\over 4}\Big(\omega_{00}+\omega_{0f}j_f
+\omega_{e0}\bar \jmath_e+\omega_{ef}\bar\jmath_ej_f\Big)
={1\over 2}\Big(\omega_{0g}-{1\over 2}\epsilon_{efg}\omega_{ef}\Big)j_g.
&\cr}
$$
Explicitly, the $\omega^\pm$ are given by the formulae
$$
\omega^+=-2\imag\Big(e^\vee e(\varphi)\Big),
\quad
\omega^-=-2\imag\Big(e(\varphi) e^\vee\Big).
\eqno(3.11)
$$
The components of the Riemann 2--form $R_{ab}$ can be assembled into the 
two quaternionic 2--forms
$$
\eqalignno{
R^+&=-{1\over 4}\Big(R_{00}+R_{0f}\bar \jmath_f
+R_{e0}j_e+R_{ef}j_e\bar \jmath_f\Big)
={1\over 2}\Big(R_{0g}+{1\over 2}\epsilon_{efg}R_{ef}\Big)j_g,
&(3.12)\cr
R^-&=+{1\over 4}\Big(R_{00}+R_{0f}j_f
+R_{e0}\bar\jmath_e+R_{ef}\bar\jmath_ej_f\Big)
={1\over 2}\Big(R_{0g}-{1\over 2}\epsilon_{efg}R_{ef}\Big)j_g.
&\cr}
$$
By explicit computation, one finds
$$
R^+=
2\imag\Big(e^\vee\wedge\big(de(\varphi)+2|e(\varphi)|^2\bar e^\vee\big)
\Big),\quad
R^-=-2\imag\Big(\big(de(\varphi)+2|e(\varphi)|^2\bar e^\vee\big)\wedge e^\vee
\Big).
\eqno(3.13)
$$
The components of the Ricci 1--form $S_a$ can be organized into 
the quaternionic 1--form
$$
S=S_0+S_ej_e.
\eqno(3.14)
$$
This is explicitly given by
$$
S=-8\Big[d\bar e(\varphi)
+2\big(\bar e(e(\varphi))+3|e(\varphi)|^2\big)e^\vee\Big].
\eqno(3.15)
$$
Finally, the Ricci scalar $s$ is given by
$$
s=-96\Big[\bar e(e(\varphi))+2|e(\varphi)|^2\Big].
\eqno(3.16)
$$

\vskip.12cm\par\noindent
{\it Proof}. We give only a sketch. For a conformally flat metric, one has
$$
\omega_{ab}=e_b(\varphi)e^\vee{}_a-e_a(\varphi)e^\vee{}_b.
\eqno(3.17)
$$
From this relation, using the standard definitions of the Riemann
2--form $R_{ab}=d\omega_{ab}+\omega_{ac}\wedge\omega_{cb}$,
the Ricci 1--form $S_a=\iota(e_b)R_{ba}$ and the
Ricci scalar $s=\iota(e_a)S_a$, it is easy to see that
$$
\eqalignno{
R_{ab}&=e^\vee{}_b\wedge Q_a-e^\vee{}_a\wedge Q_b,
&(3.18)\cr
S_a&=-2Q_a-Qe^\vee{}_a,
&(3.19)\cr
s&=-6Q,
&(3.20)\cr}
$$
where
$$
\eqalignno{
Q_a&=de_a(\varphi)+{1\over 2}e_c(\varphi)e_c(\varphi)e^\vee{}_a,
&(3.21)\cr
Q&=e_c(e_c(\varphi))+2e_c(\varphi)e_c(\varphi).
&(3.22)\cr}
$$
Using these formulae, one obtains straightforwardly the above relations.
\hfill $QED$ \vskip.12cm

From (3.11), one can derive the identity
$$
de^\vee-\omega^+\wedge e^\vee+e^\vee\wedge\omega^-=0,
\eqno(3.23)
$$
which is equivalent to the well-known relation 
$de^\vee{}_a+\omega_{ab}\wedge e^\vee{}_b=0$.
From (3.11) and (3.13), one can verify that 
$$
R^+=d\omega^+-\omega^+\wedge\omega^+, \quad
R^-=d\omega^-+\omega^-\wedge\omega^-,
\eqno(3.24)
$$
relations which are equivalent to the definition of the Riemann 2--form
$R_{ab}=d\omega_{ab}+\omega_{ac}\wedge\omega_{cb}$. Other basic relations
could be obtained in a similar manner.

Expressions of the Pontryagin density $\gamma={1\over 8\pi^2}W_{ab}\wedge 
W_{ab}$, where $W_{ab}$ is the Weyl 2--form, and of the Euler density 
$\epsilon={1\over 32\pi^2}\epsilon_{abcd}R_{ab}\wedge R_{cd}$
can similarly be obtained. For a locally conformally flat metric 
such as $g$, one obviously has
$$
\gamma=0.
\eqno(3.25)
$$
$\epsilon$ is explicitly given by
$$
\eqalignno{
\epsilon&=\Big({2\over\pi}\Big)^2\Big\{12\Big[\bar e(e(\varphi))
+2|e(\varphi)|^2\Big]^2*1
&(3.26)\cr
&\hphantom{=}-\real\Big[
\big(d\bar e(\varphi)-\bar e(e(\varphi))e^\vee\big)\wedge 
*\big(de(\varphi)-\bar e(e(\varphi))\bar e^\vee\big)\Big]\Big\}.
&\cr}
$$
\vskip.12cm\par\noindent
{\it Proof}. It is known that $\epsilon={1\over 16\pi^2}\Big\{W_{ab}\wedge 
*W_{ab}+{1\over 12}s^2 *1-\big(S_a-{1\over 4}se^\vee{}_a\big)
\wedge *\big(S_a-{1\over 4}se^\vee{}_a\big)\Big\}$. In the present case,
$W_{ab}=0$, as the metric is locally conformally flat. Using (3.19)--(3.20)
and (3.21)--(3.22), it is straightforward to derive the above formula.
\hfill $QED$ \vskip.12cm

{\it Global quaternionic Riemannian geometry of a Kulkarni 4--fold}

The quaternionic tensors constructed in the previous subsection have 
very simple covariance properties on a Kulkarni 4--fold $X$.

The matching is implemented by the $\Sp(1)$ transition functions
$$
\theta^\pm{}_{\alpha\beta}
=\eta^\pm{}_{\alpha\beta}/|\eta^\pm{}_{\alpha\beta}|,
\eqno(3.27)
$$
with $\eta^\pm{}_{\alpha\beta}$ given by (2.23).
In general, these do not form a smooth $\Sp(1)$ 1--cocycle, unless
$w=1$, as, by (2.31),
$$
\theta^\pm{}_{\alpha\gamma}
=w_{\alpha\beta\gamma}\theta^\pm{}_{\alpha\beta}\theta^\pm{}_{\beta\gamma},
\eqno(3.28)
$$
when $U_\alpha\cap U_\beta\cap U_\gamma\not=\emptyset$.
However, $\theta^\pm{}_L\otimes\theta^\mp{}_R$ and 
$\theta^\pm{}_L\otimes\theta^\pm{}_R$
are, respectively, a $(\Sp(1)\times\Sp(1))/\Bbb Z_2$ 1--cocycle
and a $\Sp(1)/\Bbb Z_2$ 1-cocycle.

We assume that the local scales $\varphi_\alpha$ match as
$$
\varphi_\alpha=\varphi_\beta-\ln|\eta^+{}_{\alpha\beta}|+
\ln|\eta^-{}_{\alpha\beta}|,
\eqno(3.29)
$$
whenever $U_\alpha\cap U_\beta\not=\emptyset$.
This is designed in such a way to 
render $g=\{g_\alpha\}$ a globally defined metric (see (3.32) below).

The matching relations for the einbein $e=\{e_\alpha\}$ and 
$e^\vee=\{e^\vee{}_\alpha\}$ are  
$$
e_\alpha=
\theta^-{}_{\alpha\beta}e_\beta
(\theta^+{}_{\alpha\beta})^{-1}
\eqno(3.30)
$$
and
$$
e^\vee{}_\alpha
=\theta^+{}_{\alpha\beta}e^\vee{}_\beta
(\theta^-{}_{\alpha\beta})^{-1},
\eqno(3.31)
$$
with $U_\alpha\cap U_\beta\not=\emptyset$.

\vskip.12cm\par\noindent
{\it Proof}. These relations follow readily from combining (2.24), (2.25)
and (3.29) with (3.5) and (3.7). \hfill $QED$ \vskip.12cm

The matching relations of the metric $g=\{g_\alpha\}$ are by construction 
$$
g_\alpha=g_\beta.
\eqno(3.32)
$$
on $U_\alpha\cap U_\beta\not=\emptyset$.
As a consequence, the Hodge star operators $*_\alpha$ associated with 
the $g_\alpha$ match as 
$$
*_\alpha=*_\beta.
\eqno(3.33)
$$

For $U_\alpha\cap U_\beta\not=\emptyset$, the matching relations 
for the spin connection 1-forms $\omega^\pm=\{\omega^\pm{}_\alpha\}$ are 
$$
\omega^\pm{}_\alpha=
\theta^\pm{}_{\alpha\beta}\omega^\pm{}_\beta
(\theta^\pm{}_{\alpha\beta})^{-1}
\pm d\theta^\pm{}_{\alpha\beta}
(\theta^\pm{}_{\alpha\beta})^{-1}.
\eqno(3.34)
$$
The matching relations for the Riemann 2--forms $R^\pm=\{R^\pm{}_\alpha\}$, 
the Ricci 1--form $S=\{S_\alpha\}$ and the Ricci scalar $s=\{s_\alpha\}$ 
are
$$
\eqalignno{
R^\pm{}_\alpha
&=\theta^\pm{}_{\alpha\beta}R^\pm{}_\beta
(\theta^\pm{}_{\alpha\beta})^{-1},
&(3.35)\cr
S_\alpha
&=\theta^+{}_{\alpha\beta}S_\beta
(\theta^-{}_{\alpha\beta})^{-1},
&(3.36)\cr
s_\alpha&
=s_\beta.
&(3.37)\cr}
$$
So, $R^\pm\in\Omega^2(X,\theta^\pm{}_L\otimes\theta^\pm{}_R)$,
$S\in\Omega^1(X,\theta^+{}_L\otimes\theta^-{}_R)$ and 
$s\in\Omega^0(X)$.

\vskip.12cm\par\noindent
{\it Proof}. The matching relation of the dual vierbein
$e^\vee{}_a=\{e^\vee_{\alpha a}\}$ is of the form
$$
e^\vee{}_{\alpha a}=r_{\alpha\beta ab}e^\vee{}_{\beta b},
\eqno(3.38)
$$
where $r_{\alpha\beta}$ is some smooth $\SO(4)$ valued function on 
$U_\alpha\cap U_\beta$. 
Combining (3.6) and (3.38) and comparing with (3.31), one finds
$$
r_{\alpha\beta 0a}+r_{\alpha\beta ea}j_e=
\theta^+{}_{\alpha\beta}\big(\delta_{0a}+\delta_{ea}j_e\big)
(\theta^-{}_{\alpha\beta})^{-1}.
\eqno(3.39)
$$
As well--known, one has 
$\omega_{\alpha ab}=r_{\alpha\beta ac}r_{\alpha\beta bd}\omega_{\beta cd}
-d r_{\alpha\beta ac}r_{\alpha\beta bc}$
and $R_{\alpha ab}=r_{\alpha\beta ac}r_{\alpha\beta bd}R_{\beta cd}$ and
$S_{\alpha a}=r_{\alpha\beta ab}S_{\beta b}$.
Using (3.39) and the definitions (3.10), (3.12) and (3.14), it is 
straightforward to check that (3.34), (3.35) and (3.36) hold. (3.37)
is obvious. \hfill $QED$ \vskip.12cm

{\it Selfduality and the Einstein condition}

Selfdual Einstein 4--folds form a broad class of Riemannian 4--folds,
which has been intensively studied \ref{14}. Consider a Kulkarni 4--fold $X$ 
equipped with the metric $g$ of the local form (3.8). 
$g$  is locally conformally flat and thus trivially selfdual. The Einstein 
condition, conversely, is non trivial.

The metric $g$ is Einstein if and only if, locally,
$$
d\bar e(\varphi)-\bar e(e(\varphi))e^\vee=0.
\eqno(3.40)
$$
The local solution of this equation is 
$$
\ee^{-\varphi}=w+2\real(\bar vq)+u|q|^2, 
\quad \hbox{with $u$, $w\in\Bbb R$, $v\in \Bbb H$}.
\eqno(3.41)
$$

\vskip.12cm\par\noindent
{\it Proof}. The Einstein condition states that
$S_a-(s/4)e^\vee{}_a=0$.
Using the definitions (3.6) and (3.14) and the formulae (3.15) and (3.16),
one gets readily (3.40). Explicitly, using (3.5) and (3.7),
(3.40) can be cast as 
$$
d(\partial_{\bar q}\ee^{-\varphi})
-dq\partial_{qL}(\partial_{\bar q}\ee^{-\varphi})=0.
\eqno(3.42)
$$
Now, for any smooth $\Bbb H$--valued 
function $f$, the condition $df-dq\partial_{qL}f=0$
restricts $f$ to be of the form $f(q)=a+qb$ with $a,~b\in\Bbb H$
\ref{7}. Hence, (3.42) entails that
$$
\partial_{\bar q}\ee^{-\varphi}=(v+qu)/2,
\quad {\rm with}~~u,~v\in \Bbb H.
\eqno(3.43)
$$
From (3.43), using that $\ee^{-\varphi}$ is real valued, one gets
$$
d\ee^{-\varphi}=dq(\bar v+\bar u\bar q)+(v+qu)d\bar q.
\eqno(3.44)
$$
The integrability condition $d^2\ee^{-\varphi}=0$ yields the 
equation $dq\wedge(u-\bar u)d\bar q=0$, which, as is easy to see, 
entails that $u-\bar u=0$ or $u\in\Bbb R$. So,
$$
d\ee^{-\varphi}=d\Big[2\real(\bar vq)+u|q|^2\Big],
\eqno(3.45)
$$
which, upon integration, yields (3.41). \hfill $QED$ \vskip.12cm

For $U_\alpha\cap U_\beta\not=\emptyset$, we set 
$$
K_{\alpha\beta}
=(|\eta^+{}_{\alpha\beta}||\eta^-{}_{\alpha\beta}|)^{-{1\over 2}}
T_{\alpha\beta},
\eqno(3.46)
$$
with $T_{\alpha\beta}$ defined in (2.22) and 
$\eta^\pm{}_{\alpha\beta}$ given by (2.23).
Then, $K_{\alpha\beta}$ does not depend on the choice of representative 
of $T_{\alpha\beta}\in\PGL(2,\Bbb H)$ in $\GL(2,\Bbb H)$. Further, 
$K=\{K_{\alpha\beta}\}$ is a flat $\GL(2,\Bbb H)$ 1--cochain
satisfying relation (2.30) with $T_{\alpha\beta}$ substituted by 
$K_{\alpha\beta}$. For an Einstein metric of the form (3.41), set
$$
M=\bigg(\matrix{u &v \cr\bar v & w \cr}\bigg).
\eqno(3.47)
$$
Then, one has the matching relation 
$$
M_\beta=K_{\alpha\beta}{}^\dagger M_\alpha K_{\alpha\beta}.
\eqno(3.48)
$$

\vskip.12cm\par\noindent
{\it Proof}. In the proof of relation (2.30), it was shown that
$|\eta^+{}_{\alpha\beta}||\eta^-{}_{\alpha\beta}|$ is a positive constant.
Using this fact (2.30) and (2.31), 
it is immediate to see that $K=\{K_{\alpha\beta}\}$ is a 
flat $\GL(2,\Bbb H)$ 1--cochain satisfying (2.30). Independence from
choices of representative is evident from the definition (3.46) and from 
(2.23). The above matching relation follows from (3.29), upon writing
$$
\ee^{-\varphi}=(\bar q,1)\Big(\matrix{u &v \cr\bar v & w \cr}\Big)
\Big(\matrix{q \cr 1\cr}\Big)
\eqno(3.49)
$$ 
and using (2.22), (1.11) and (2.23). \hfill $QED$ \vskip.12cm

\par\noindent
This result is interesting. It reduces the problem of finding 
a locally conformally flat Einstein metric to the problem of finding 
a flat positive definite section $M=\{M_\alpha\}$ of the flat 1--cocycle
${\rm Sq}K$, where, for any $A\in\GL(2,\Bbb H)$, ${\rm Sq}AU=A^\dagger UA$,
for $U$ a 2 by 2 matrix on $\Bbb H$.

{\it The conformal d'Alembertian} $W$ {\it and the d'Alembertian}
$\square$

Let $X$ be a Kulkarni 4--fold with the metric $g$ of eq. (3.8).
The conformal d'A\-lem\-ber\-tian $W$ of $g$ is defined by
$$
Wf=d*df-{s\over 6}f*1, 
\eqno(3.50)
$$
for $f\in\Omega^0(X)$. So, $Wf\in\Omega^4(X)$.
$W$ is simply related to the operator $\square$ is defined in (2.39).
Indeed, $\ee^\varphi f\in\Omega^0(X,\rho)$ and 
$$
Wf=16\ee^\varphi\square(\ee^\varphi f).
\eqno(3.51)
$$

\vskip.12cm\par\noindent
{\it Proof}. Combining (2.38) and (3.29), one verifies easily that
$\ee^\varphi f\in\Omega^0(X,\rho)$ if $f\in\Omega^0(X)$.
As is well--known, the operator $W$ is conformally covariant.
If $g_0$ and $g=\ee^h g_0$ are two conformally related metrics, then
$Wf=\ee^h W_0(\ee^h f)$. If we take $g_0$ to be the flat 
metric and $g$ to be the metric (3.3), we get (3.51) 
readily. \hfill $QED$ \vskip.12cm

\par\noindent
(3.51) entails immediately an isomorphism $\ker W\cong\ker\square$ 
of $\Bbb R$ linear spaces. 

A well-known argument \`a la Bochner shows that, if $X$ is compact and 
$s\geq 0$ and $s\not\equiv 0$ on $X$, then $\dim\ker W=0$. So, on a compact 
Kulkarni 4--fold $X$ such that the associated conformal class of 
locally conformally flat metrics contains a representative 
whose $s$ has the above properties, $\dim\ker\square=0$, that is 
there are no harmonic $F\in\Omega^0(X,\rho)$.

{\it The Dirac operator} $/\!\!\!\!D$ {\it and the Fueter operators}
$\bar\partial_{R,L}$

Let $X$ be a Kulkarni 4--fold with $w=1$ equipped with the metric $g$ of
eq. (3.8). We set $\sigma^+=\theta^+{}_R$ and $\sigma^-=\theta^-{}_L$. 
Owing to (3.28), as $w=1$, the $\sigma^\pm$ are smooth $\Sp(1)$
1--cocycles depending on a choice of a flat $\Bbb Z_2$ 1--cocycle $a$.
We set $\sigma=\sigma^+\oplus\sigma^-$. So, any 
$\lambda\in\Omega^0(X,\sigma)$ 
is of the form $\lambda=\lambda^+\oplus\lambda^-$ with 
$\lambda^\pm\in\Omega^0(X,\sigma^\pm)$. We set 
$$
(\lambda_1,\lambda_2)
=\real(\overline{\lambda_1{}^+}\lambda_2{}^+)
+\real(\overline{\lambda_1{}^-}\lambda_2{}^-),
\eqno(3.52)
$$
for $\lambda_1,~\lambda_2\in\Omega^0(X,\sigma)$ and,  
for a vector field $u$ on $X$,
$$
/\!\!\!u\lambda=(\overline{\lambda^-}\langle\bar e^\vee, u\rangle)\oplus
(\langle\bar e^\vee, u\rangle\overline{\lambda^+}),
\eqno(3.53)
$$
for $\lambda\in\Omega^0(X,\sigma)$.
Then, $\Omega^0(X,\sigma)$ is a real Clifford module on $(X,g)$ with 
Clifford inner product and Clifford action given respectively by
(3.52) and (3.53).

\vskip.12cm\par\noindent
{\it Proof}. If $\lambda^\pm\in\Omega^0(X,\sigma^\pm)$, one has
$\lambda^+{}_\alpha=\lambda^+{}_\beta\theta^+{}_{\beta\alpha}$ and
$\lambda^-{}_\alpha=\theta^-{}_{\alpha\beta}\lambda^-{}_\beta$,
whenever defined. Further, $|\theta^\pm{}_{\alpha\beta}|=1$, by (3.27).
Taking these relations into account, one verifies that 
$(\lambda_1,\lambda_2)_\alpha=(\lambda_1,\lambda_2)_\beta$.
So, the Clifford inner product is well-defined. Using 
the same relations once more and (3.31), one verifies also that 
$(/\!\!\!u\lambda)^+{}_\alpha=(/\!\!\!u\lambda)^+{}_\beta
\theta^+{}_{\beta\alpha}$ and $(/\!\!\!u\lambda)^-{}_\alpha
=\theta^-{}_{\alpha\beta}(/\!\!\!u\lambda)^-{}_\beta$. So, $/\!\!\!u$ 
maps linearly $\Omega^0(X,\sigma^\pm)$ into $\Omega^0(X,\sigma^\mp)$. 
Finally, one checks easily that $(\lambda_1,/\!\!\!u\lambda_2)
=(/\!\!\!u\lambda_1,\lambda_2)$ and, by using (3.8), that 
$/\!\!\!u^2=g(u,u)1$.  \hfill $QED$ \vskip.12cm

For $\lambda\in\Omega^0(X,\sigma)$, we define 
$$
D\lambda
=\Big(d\lambda^++\lambda^+\omega^+\Big)
\oplus\Big(d\lambda^-+\omega^-\lambda^-\Big).
\eqno(3.54)
$$
Then, $D$ is a Clifford connection for the Clifford 
module $\Omega^0(X,\sigma)$.

\vskip.12cm\par\noindent
{\it Proof}. Using (3.34) and the matching relations of $\lambda^\pm$
given above, it is straightforward to check that 
$(D\lambda)^+{}_\alpha=(D\lambda)^+{}_\beta\theta^+{}_{\beta\alpha}$ and
$(D\lambda)^-{}_\alpha=\theta^-{}_{\alpha\beta}(D\lambda)^-{}_\beta$,
whenever defined. So, $D$ maps $\Omega^0(X,\sigma^\pm)$ into
$\Omega^1(X,\sigma^\pm)$. $D$ manifestly has the properties 
defining a connection on $\Omega^0(X,\sigma)$. From the identity
$\nabla_ve^\vee{}_a+\langle\omega_{ab},v\rangle e^\vee{}_b=0$,
where $\nabla$ is the Levi--Civita connection and $v$ a vector field on $X$, 
and from (3.6) and (3.10), it is straightforward to show that 
$\nabla_ve^\vee-\langle\omega^+,u\rangle e^\vee
-e^\vee\langle\omega^-,u\rangle=0$. Using this latter identity,
one checks by simply applying the definitions (3.53) and (3.54) that 
$[D,/\!\!\!u]=\nabla/\!\!\!u$. This shows that $D$ is a Clifford connection.
\hfill $QED$ \vskip.12cm

The Dirac operator $/\!\!\!\!D$ associated with the Clifford connection
$D$ of the Clifford module $\Omega^0(X,\sigma)$ is readily obtained:
$$
/\!\!\!\!D\lambda
=\Big(4\langle\overline{(D\lambda)^-},e\rangle\Big)\oplus
\Big(4\langle e,\overline{(D\lambda)^+}\rangle\Big),
\eqno(3.55)
$$
with $\lambda\in\Omega^0(X,\sigma)$.
This is very simply related to the Fueter operators 
$\bar\partial_{R,L}$ defined in (2.44).
Indeed, $\ee^{{3\over 2}\varphi}\lambda^\pm\in\Omega^0(X,\varpi^\pm)$
and 
$$
/\!\!\!\!D\lambda=\Big(4\overline{\ee^{-{5\over 2}\varphi}
\partial_{\bar qL}(\ee^{{3\over 2}\varphi}\lambda^-)}\Big)\oplus
\Big(4\overline{(\lambda^+\ee^{{3\over 2}\varphi})\partial_{\bar qR}
\ee^{-{5\over 2}\varphi}}\Big).
\eqno(3.56)
$$

\vskip.12cm\par\noindent
{\it Proof}. Combining (2.43), (3.27), (3.29) and the matching 
relations of the $\lambda^\pm$, is easily seen that 
$\ee^{{3\over 2}\varphi}\lambda^\pm\in\Omega^0(X,\varpi^\pm)$.
From (3.1), (3.2) and (3.17), one has that
$\omega_{ab}=\delta_{ai}\delta_b{}^j\partial_{xj}\varphi dx^i
-\delta_{bj}\delta_a{}^i\partial_{xi}\varphi dx^j$. 
Using this relation, (3.5) and (2.2), it is easy to verify that 
$\langle (D\lambda)^+,\bar e\rangle
=(\lambda^+\partial_{\bar qR}+{3\over 2}\lambda^+\varphi\partial_{\bar qL})
\ee^{-\varphi}$ and 
$\langle \bar e,(D\lambda)^-\rangle
=\ee^{-\varphi}(\partial_{\bar qL}\lambda^-
+{3\over 2}\partial_{\bar qL}\varphi\lambda^-)$.
Using these expressions in (3.55), one gets (3.56) immediately. 
\hfill $QED$ \vskip.12cm

\par\noindent 
It follows immediately from (3.56) that $\ker\bar\partial_R
\cong\ker/\!\!\!\!D|_{\Omega^0(X,\sigma^+)}$ 
and $\ker\bar\partial_L
\cong\ker/\!\!\!\!D|_{\Omega^0(X,\sigma^-)}$, where the first (second)
isomorphism is left (right) $\Bbb H$--linear.

The Dirac operator $/\!\!\!\!D$ satisfies the well-known 
Bochner--Lichnerowicz--Weitzenboek formula 
$/\!\!\!\!D^2=-\square_D+{1\over 4}s$, with $\square_D$ the d'Alembertian of 
the Clifford connection $D$. By a well-known argument \`a la Bochner,
we see that, if $X$ is compact and $s\geq 0$ and $s\not\equiv 0$ on $X$, 
then $\dim\ker/\!\!\!\!D=0$. So, on a compact Kulkarni 4--fold $X$ such that 
the associated conformal class of locally conformally 
flat metrics contains a representative whose $s$ has the above properties, 
$\dim\ker\bar\partial_{R,L}=0$, that is there are no Fueter holomorphic
$\Phi\in\Omega^0(X,\varpi^+)$ and $\Psi\in\Omega^0(X,\varpi^-)$.

When $X$ is compact, one can compute the index of $/\!\!\!\!D$,
$\ind/\!\!\!\!D$, by using the Atiyah--Singer index theorem. One has 
$$
\ind/\!\!\!\!D=\dim\ker/\!\!\!\!D|_{\Omega^0(X,\sigma^+)}-
\dim\ker/\!\!\!\!D|_{\Omega^0(X,\sigma^-)}=0
\eqno(3.57)
$$

\vskip.12cm\par\noindent 
{\it Proof}. Using (3.53) and (3.54) and taking (3.24) into account, 
one finds that $D^2\lambda-{1\over 4}R_{ab}/\!\!\!e_a/\!\!\!e_b\lambda=
\big(\lambda^+(d\omega^+-\omega^+\wedge\omega^+
-R^+)\big)\oplus\big((d\omega^-+\omega^-\wedge\omega^--R^-)\lambda^-\big)
=0$. The Clifford connection $D$ has thus no twisting. 
In this case, the Atiyah--Singer index theorem gives
$\ind/\!\!\!\!D=-{1\over24}\int_Xp_1(X)$. On 
account of (2.54), $\ind/\!\!\!\!D=0$. \hfill $QED$ \vskip.12cm

\par\noindent 
When $X$ is compact, we conclude from (3.57) that 
$$
\dim\ker\bar\partial_R=\dim\ker\bar\partial_L.
\eqno(3.58)
$$
The number of right Fueter holomorphic sections 
$\Phi\in\Omega^0(X,\varpi^+)$ equals the number of left
Fueter holomorphic sections $\Psi\in\Omega^0(X,\varpi^-)$.

{\it The isometry group of the metric} $g$

Given a metric $g$ on $X$ of the form (3.8), we denote by ${\rm UAut}(X,g)$ 
the subgroup of $\Aut(X)$ leaving $g$ invariant. 

{\it Examples of special metrics}

Below, we shall consider the Kulkarni 4--folds of Kleinian type  
$\Gamma\backslash D$, which were described at the end of section 2.

$i)$ $D=\Bbb H\Bbb P^1$. $\Bbb H\Bbb P^1$ has the distinguished metric
$$
g={4\real(d\bar q\otimes dq)\over(1+|q|^2)^2}.
\eqno(3.59)
$$
$g$ is nothing but the customary round metric of $S^4$. As is well-known,
$g$ is Einstein with $s=12$. 
${\rm UAut}(\Bbb H\Bbb P^1,g)$ is the subgroup of $\PGL(2,\Bbb H)$ formed 
by those $T$ such that $|T_{11}|^2+|T_{21}|^2=|T_{22}|^2+|T_{12}|^2=k$ for 
some $k\in\Bbb R_+$ and $\bar T_{11}T_{12}+\bar T_{21}T_{22}=0$ and is thus
a proper subgroup of $\Aut(\Bbb H\Bbb P^1)$.

$ii)$ $D=\Bbb H^1$. $\Bbb H^1$ has the distinguished metric
$$
g=4\real(d\bar q\otimes dq)
\eqno(3.60)
$$
So, $g$ is the flat euclidean metric of $\Bbb R^4$. 
${\rm UAut}(\Bbb H^1,g)$ is the subgroup of $\PGL(2,\Bbb H)$ formed 
by those $T$ such that $|T_{11}|=|T_{22}|=1$ and $T_{21}=0$ and is thus
a proper subgroup of $\Aut(\Bbb H^1)$. This metric induces a special metric
on each Kulkarni 4--folds $\Gamma\backslash\Bbb H^1$ since, as it is easy 
to show, every Kleinian group $\Gamma$ for $\Bbb H^1$ is contained in 
${\rm UAut}(\Bbb H^1,g)$.

$iii)$ $D=B_1(\Bbb H^1)$. $B_1(\Bbb H^1)$ has the distinguished metric
$$
g={4\real(d\bar q\otimes dq)\over(1-|q|^2)^2}.
\eqno(3.61)
$$
As appears, $g$ is nothing but the Poincar\'e metric of $B_1(\Bbb R^4)$. 
$g$ is Einstein with $s=-12$. One checks that ${\rm UAut}(B_1(\Bbb H^1),g)$ 
is the subgroup of $\PGL(2,\Bbb H)$ formed by those 
$T$ such that $|T_{11}|^2-|T_{21}|^2=|T_{22}|^2-|T_{12}|^2=k$ for some
$k\in\Bbb R_+$ and $\bar T_{11}T_{12}-\bar T_{21}T_{22}=0$, so that 
${\rm UAut}(B_1(\Bbb H^1),g)=\Aut(B_1(\Bbb H^1))$. Therefore, this metric 
induces a special metric on each Kulkarni 4--folds 
$\Gamma\backslash B_1(\Bbb H^1)$ for every Kleinian group $\Gamma$ for 
$B_1(\Bbb H^1)$.

$iv)$ $D=\Bbb H^1-\{0\}$. $\Bbb H^1-\{0\}$ has the 
special metric
$$
g={\real(d\bar q\otimes dq)\over |q|^2}.
\eqno(3.62)
$$
One can show that ${\rm UAut}(\Bbb H^1-\{0\},g)
=\Aut(\Bbb H^1-\{0\})$ \ref{9}. Therefore, this metric 
induces a special metric on each Kulkarni 4--folds 
$\Gamma\backslash(\Bbb H^1-\{0\})$ for every Kleinian group 
$\Gamma$ for $\Bbb H^1-\{0\}$. 

$v)$ $D=\Bbb H^1-\Bbb R^1$. $\Bbb H^1-\Bbb R^1$ has the 
special metric
$$
g={\real(d\bar q\otimes dq)\over |\imag q|^2}.
\eqno(3.63)
$$
It is possible to show that ${\rm UAut}(\Bbb H^1-\Bbb R^1)
=\Aut(\Bbb H^1-\Bbb R^1)$ \ref{9}. Therefore, this metric 
induces a special metric on each Kulkarni 4--folds 
$\Gamma\backslash(\Bbb H^1-\Bbb R^1)$ for every Kleinian group 
$\Gamma$ for $\Bbb H^1-\Bbb R^1$. 
\par\vskip.6cm
\item{\bf 4.} {\bf Classical 4--dimensional conformal
field theory and Kulkarni geometry}
\vskip.4cm
\par
In this section, we consider first some general properties
of a classical conformal field theory on a Kulkarni 4--fold $X$.
Later, we illustrate two basic models, the complex scalar and the
Dirac fermion (see \ref{15} for background).

Below, we shall assume that $X$ is compact. In this way, 
integrals are convergent and, as $X$ has no boundary (see sect. 2), 
integration by parts can be carried out without picking boundary 
contributions.

{\it The classical action}

The classical action of a conformal field theory on a 4--fold $X$ is some 
local functional ${\cal I}(\Phi,e^\vee)$ of a set of conformal fields $\Phi$
and a dual vierbein $e^\vee{}_a$. By conformal invariance, for any smooth 
function $f$ on $X$, one has
$$
{\cal I}(\ee^{-f\Lambda}\Phi,\ee^fe^\vee)={\cal I}(\Phi,e^\vee),
\eqno(4.1)
$$
where $\Lambda$ is the matrix of the conformal weights of the fields $\Phi$.

Consider now a conformally flat background $e^\vee{}_a$ of the form (3.2). 
Because of conformal invariance, one has that
$$
{\cal I}(\Phi,e^\vee)=I(\phi),
\eqno(4.2)
$$
where 
$$
\phi=\ee^{\varphi \Lambda}\Phi
\eqno(4.3)
$$
is a conformally invariant field. The functional $I(\phi)$ 
depends only on $\phi$ and the underlying conformal structure.

As $\varphi$ is defined only locally and the local representations
match as in (3.29), the matching relations of 
the local representations of $\phi$ are different from 
those of the local representations of $\Phi$.
On a Kulkarni 4--fold $X$, $\phi$ is a section of some vector bundle
constructed from the $\eta^\pm$ such as $\rho$ and $\varpi^\pm$.

{\it The energy--momentum tensor}

In a classical field theory on a 4--fold $X$, 
the energy--momentum tensor is the 1--form ${\cal T}_a(\Phi,e^\vee)$, 
$a=0$, 1, 2, 3,
valued in the orthonormal frame bundle, defined by the variational identity
$\delta_{e^\vee}{\cal I}
=-{1\over 2\pi^2}\int_X\langle {\cal T}_a,\delta e_a\rangle *1$,
where $\delta_{e^\vee}\Phi=-{1\over 4}\Lambda\delta\ln e\Phi$ with
$e=\det e^\vee$ \ref{15}.
If the field theory is conformal, the energy--momentum tensor is traceless 
and thus satisfies
$$
\iota(e_a){\cal T}_a=0.
\eqno(4.4)
$$
The invariance of the classical action $\cal I$ under the action of the 
group of the automorphisms of the orthonormal frame bundle implies that, 
for classical field configurations solving the classical field equations,
the energy--momentum tensor is symmetric and conserved \ref{15}. 
The symmetry is encoded in the relation
$$
{\cal T}_a\wedge e^\vee{}_a=0.
\eqno(4.5)
$$
The conservation equation can be cast as
$$
d*{\cal T}_a+\omega_{ab}\wedge *{\cal T}_b=0.
\eqno(4.6)
$$

For a classical conformal field theory, one has
$$
{\cal T}_a(\ee^{-f\Lambda}\Phi,\ee^fe^\vee)=\ee^{-3f}{\cal T}_a(\Phi,e^\vee),
\eqno(4.7)
$$
for any smooth function $f$. This is an immediate consequence of the 
conformal invariance of the action (eq. (4.1)) and of the definition of 
${\cal T}_a$. Consequently, in a locally conformally flat metric 
background $e^\vee{}_a$ of the form (3.2), one has that
$$
{\cal T}_a(\Phi,e^\vee)=\delta_{ai}\ee^{-3\varphi}T_i(\phi),
\eqno(4.8)
$$
where the $T_i(\phi)$, $i=0$, 1, 2, 3 are 1--forms depending only on $\phi$ 
and the underlying conformal structure. They can be assembled into
the quaternionic field
$$
T={1\over 4}(T_0-T_rj_r).
\eqno(4.9)
$$
Then, it is simple to verify that the tracelessness relation (4.4) takes 
the form
$$
\real\big(T\iota(\partial_{\bar q})\big)=0.
\eqno(4.10)
$$
For classical field configurations, the symmetry relation (4.5) reads as
$$
\real\big(dq\wedge T)=0,
\eqno(4.11)
$$
while, more importantly, the conservation equation (4.6) becomes simply
$$
d\star T=0.
\eqno(4.12)
$$
This equation no longer contains any explicit dependence on the scale 
$\varphi$ of the metric background. Its validity depends crucially on 
the tracelessness and symmetry relations (4.10) and (4.11).

\vskip.12cm\par\noindent
{\it Proof}. (4.10) and (4.11) are trivial consequences of (4.4) and (4.5)
following from (4.9), (3.1), (3.2), (3.5) and (3.7).
(4.12) follows from substituting (3.9), (3.17) and (4.8) 
into (4.6) upon using (3.1)--(3.2) and (4.4)--(4.5). 
\hfill $QED$ \vskip.12cm

On a Kulkarni 4--fold $X$, $T\in\Omega^1(X,\zeta_3)$, where 
$\zeta_3$ is given by (2.34).

\vskip.12cm\par\noindent
{\it Proof}. By (4.8) and (4.9), one has  
$$
T_\alpha
={\ee^{3\varphi_\alpha}\over 4}
\Big({\cal T}_{\alpha 0}-{\cal T}_{\alpha e}j_e\Big).
\eqno(4.13)
$$
Now, on $U_\alpha\cap U_\beta\not=\emptyset$, one has 
$$
{\cal T}_{\alpha a}=r_{\alpha\beta ab}{\cal T}_{\beta b},
\eqno(4.14)
$$
where $r_{\alpha\beta}$ is the same $\SO(4)$ valued 
function as that appearing  
in (3.38). Combining (3.27), (3.29), (3.39) and (4.14) and 
recalling (2.34), 
one checks easily that the matching relation of the $T_\alpha$ is the 
required one. \hfill $QED$ \vskip.12cm

\par\noindent
In general, for an object of the same tensor 
type as $T$, the conservation equation
(4.12) would not be covariant. In the present case, it is thanks to the 
tracelessness and symmetry properties (4.10)--(4.11).

{\it The} ${\rm U}(1)$ {\it current}

In a classical field theory with a ${\rm U(1)}$ symmetry, the ${\rm U}(1)$ 
current is the 1--form ${\cal J}(\Phi,e^\vee)$ defined by 
the variational condition $\delta_\Phi{\cal I}|_{\delta\Phi=if\Phi}
=-{1\over 2\pi^2}\int_X{\cal J}\wedge *df$ 
for any function $f$ \ref{15}. For classical field
configurations solving the classical field equations, $\cal J$ satisfies
the conservation equation
$$
d*{\cal J}=0.
\eqno(4.15)
$$

For a classical conformal field theory, one has
$$
{\cal J}(\ee^{-f\Lambda}\Phi,\ee^fe^\vee)=\ee^{-2f}{\cal J}(\Phi,e^\vee),
\eqno(4.16)
$$
for any smooth function $f$. This is an immediate consequence of the 
conformal invariance of the action (eq. (4.1)) and of the definition of 
${\cal J}$. In the locally conformally flat metric 
background $e^\vee{}_a$ of eq. (3.2), one has then
$$
{\cal J}(\Phi,e^\vee)=\ee^{-2\varphi}J(\phi),
\eqno(4.17)
$$
where $J(\phi)$ is a 1--form depending only on $\phi$ and the underlying 
conformal structure. The conservation equation (4.15) takes then the form
$$
d\star J=0.
\eqno(4.18)
$$

\vskip.12cm\par\noindent
{\it Proof}. This follows readily from (4.15) upon combining
(3.9) and (4.17). \hfill $QED$ \vskip.12cm

\par\noindent
This equation no longer contains any explicit dependence on the scale 
$\varphi$ of the metric background. 

If $X$ is a Kulkarni 4--fold, $J\in\Omega^1(X,\rho^2)$, 
where $\rho$ is defined in (2.38).

\vskip.12cm\par\noindent
{\it Proof}. Immediate from (3.29) and (4.17). \hfill $QED$ \vskip.12cm

\par\noindent
Then, by (2.29), $\star J\in\Omega^3(X)$. The conservation 
equation (4.18) is thus manifestly covariant.

{\it The biquaternion algebra}

The models examined below involve the complexification of the 
quaternion field $\Bbb H$, the complex biquaternion algebra 
$\Bbb H\otimes\Bbb C$. In this brief algebraic interlude, 
we recall a few basic facts about $\Bbb H\otimes\Bbb C$
and introduce basic notation.

Here and below, to avoid possible confusion with 
the corresponding quaternionic operations, we denote complex conjugation  
by $\bar{\phantom{\Phi}}_{\rm c}$ and complex real (imaginary) part by  
$\real_{\rm c}$ ($\imag_{\rm c}$).

A generic element $z\in\Bbb H\otimes\Bbb C$ can be represented
as a real linear combination of elements of the form $a\otimes\zeta$, 
where $a\in\Bbb H$ and $\zeta\in\Bbb C$. As a complex algebra,
$\Bbb H\otimes\Bbb C$ carries a conjugation $\bar{\phantom{z}}$ 
defined by $\overline{a\otimes\zeta}=\bar a\otimes\bar\zeta_{\rm c}$ and an 
antilinear involution $\tilde{\phantom{z}}$ defined by 
$\widetilde{a\otimes\zeta}=a\otimes\bar\zeta_{\rm c}$
\footnote{}{}
\footnote{${}^5$}{A conjugation (antilinear involution) $K$ on a complex 
algebra $A$ is an antilinear map $K:A\to A$ such that $K^2=1_A$ and that, 
for $a,~b\in A$, $K(ab)=K(b)K(a)$ ($K(ab)=K(a)K(b))$.}.
$\Bbb H$ can be canonically identified with the subalgebra
of $\Bbb H\otimes\Bbb C$ fixed by $\tilde{\phantom{z}}$.
The action of the conjugation $\bar{\phantom{z}}$ on this subalgebra 
coincides with the quaternionic conjugation $\bar{\phantom{z}}$ 
as defined earlier.

There is a canonical algebra isomorphism $c:\Bbb C(2)\to
\Bbb H\otimes\Bbb C$, where $\Bbb C(2)$ is the 
complex algebra of 2 by 2 complex matrices.
Denoting by $\tau_f$, $f=1$, 2, 3, $-i$ times the standard Pauli
matrices, $c$ is uniquely defined by $c(1_2)=1\otimes 1$ and 
$c(\tau_f)=j_f\otimes 1$. The isomorphism $c$ has the properties that 
$\det M=\tilde{\bar c}(M)c(M)$ 
and that $c(M^\dagger_{\rm c})
=\bar c(M)$ and $c(C^{-1}\bar M_{\rm c}C)=\tilde c(M)$ for any 
$M\in\Bbb C(2)$, where $C=\Big(\matrix{\hphantom{-}0&1\cr-1&0\cr}\Big)$ is 
the conjugation matrix.

{\it The complex scalar}

Consider a complex scalar field $\Phi$ with action
$$
{\cal I}(\Phi,\bar\Phi_{\rm c},e^\vee)=
{1\over 2\pi^2}
\int_Xd^4xg^{1\over 2}\Big[g^{ij}\partial_i\bar\Phi_{\rm c}\partial_j\Phi
+{1\over 6}s\bar\Phi_{\rm c}\Phi\Big],
\eqno(4.19)
$$
where $g$ is the metric corresponding to $e^\vee{}_a$ and $s$ is the Ricci 
scalar. The field $\Phi$ has conformal weight $\Lambda=1$. It is well-known 
that the above action is conformally invariant \ref{15}.

The conformally invariant field $\phi$ corresponding to $\Phi$ is thus 
given by
$$
\phi=\ee^\varphi\Phi.
\eqno(4.20)
$$
Then, $\phi\in\Omega^0(X,\rho)\otimes\Bbb C$, where $\rho$ is defined 
in (2.38). 

\vskip.12cm\par\noindent
{\it Proof}. By (3.29). \hfill $QED$ \vskip.12cm

In terms of $\phi$, the action functional is simply
$$
I(\phi,\bar\phi_{\rm c})=-{8\over \pi^2}\int_X\bar\phi_{\rm c}\square\phi,
\eqno(4.21)
$$
where $\square$ is defined in (2.39).
The integrand belongs to $\Omega^4(X)$, as 
$\square\phi\in\Omega^0(X,\rho^{-1})\otimes\Bbb C$,
and integration is thus well-defined. 

\vskip.12cm\par\noindent
{\it Proof}. This follows from substituting (3.20) and (3.22), 
upon using (3.1), and (4.20) into (4.19), by a 
straightforward calculation. \hfill $QED$ \vskip.12cm

The classical field equations of $\Phi$ are \ref{15}
$$
\nabla^j\nabla_j\Phi-{1\over 6}s\Phi=0.
\eqno(4.22)
$$
In terms of the field $\phi$, they read simply as
$$
\square\phi=0,
\eqno(4.23)
$$
that is $\phi$ is harmonic. See the discussion of section 3 concerning 
the solutions of this equation. 

The energy--momentum tensor of the complex scalar $\Phi$ 
is given by \ref{15} 
$$
\eqalignno{
{\cal T}_a(\Phi,\bar\Phi_{\rm c},e^\vee)
&=\real_{\rm c}\Big\{{2\over 3}\Big[
\bar\Phi_{\rm c} e_a{}^j\nabla_j\nabla_i\Phi
-{1\over 4}\bar\Phi_{\rm c}\nabla^k\nabla_k\Phi e^\vee{}_{ai}\Big]
&(4.24)\cr
&\hphantom{=}
-{4\over 3}\Big[e_a{}^j\nabla_j\bar\Phi_{\rm c}\nabla_i\Phi
-{1\over 4}\nabla^k\bar\Phi_{\rm c}\nabla_k\Phi e^\vee{}_{ai}\Big]
-{1\over 3}\Big[S_{ai}-{1\over 4}se^\vee{}_{ai}\Big]
\bar\Phi_{\rm c}\Phi\Big\}dx^i.
&\cr}
$$
One can verify that (4.8) holds.  
The conformally invariant energy--momentum tensor $T$ is given by
$$
\eqalignno{
T(\phi,\bar\phi_{\rm c})
&=-{2\over 3}\Big\{\partial_q\bar\phi_{\rm c} d\phi
-{1\over 2}\bar\phi_{\rm c} d\partial_q\phi
+\partial_q\phi d\bar\phi_{\rm c}
-{1\over 2}\phi d\partial_q\bar\phi_{\rm c}
&(4.25)\cr
&\hphantom{=}-\Big(\partial_{\bar q}\bar\phi_{\rm c}\partial_q\phi
-{1\over 2}\bar\phi_{\rm c}\partial_{\bar q}\partial_q\phi
+\partial_{\bar q}\phi\partial_q\bar\phi_{\rm c}
-{1\over 2}\phi\partial_{\bar q}\partial_q\bar\phi_{\rm c}\Big)d\bar q\Big\}.
&\cr}
$$
We have checked that $T$ satisfies (4.10) and that (4.11) and (4.12) hold, 
when $\phi$ fulfills the field equations (4.23). For a  
field configuration $\phi$ satisfying (4.23), 
the second and fourth term proportional to $d\bar q$ 
in (4.25) are zero.

The model considered has an obvious ${\rm U}(1)$ symmetry. The 
corresponding ${\rm U}(1)$ current is
$$
{\cal J}(\Phi,\bar\Phi_{\rm c},e^\vee)
=2\imag_{\rm c}\big(\Phi\partial_i\bar\Phi_{\rm c}\big)dx^i.
\eqno(4.26)
$$
It is easy to see that (4.17) is fulfilled with
$$
J(\phi,\bar\phi_{\rm c})
={1\over i}\big(\phi d\bar\phi_{\rm c}-\bar\phi_{\rm c} d\phi\big).
\eqno(4.27)
$$
One verifies readily that $J$ satisfies (4.18), when $\phi$ satisfies 
the field equations (4.23).

{\it The Dirac fermion}

Suppose that $w=1$, so that $X$ is spin, and let us fix the spin structure.
Consider a euclidean Dirac fermion field $\Psi$. $\Psi\in\Pi
\Omega^0(X,\Sigma^+\oplus\Sigma^-)$, where 
$\Sigma^\pm$ are the positive/negative chirality spinor bundles and the 
notation $\Pi V$ indicates the Grassmann odd partner of a vector 
space $V$.

The Dirac action is 
$$
{\cal I}(\Psi,\Psi^\dagger_{\rm c},e^\vee)=
{1\over 2\pi^2}
\int_Xd^4xei\Psi^\dagger_{\rm c}\gamma_ae_a{}^jD_j\Psi,
\eqno(4.28)
$$
where $D$ is the spin covariant derivative, 
$D_j\Psi=(\partial_j+{1\over 4}\omega_{abj}\gamma_a\gamma_b)\Psi$,
the $\gamma_a$, $a=0$, 1, 2, 3, being the euclidean gamma
matrices satisfying $\gamma_a\gamma_b+\gamma_b\gamma_a=2\delta_{ab}$
and $\gamma_a{}^\dagger_{\rm c}=\gamma_a$. 
The field $\Psi$ has conformal weight
$\Lambda=3/2$. As is well-known, the above action is conformally 
invariant \ref{15}.
We shall write the action in a way such that
its connection with the underlying Kulkarni geometry becomes 
manifest. 

Fix $v_0\in\Bbb C^2$, $v_0\not=0$. We define a linear map
$Q:\Bbb C^2\to\Bbb H\otimes\Bbb C$ by
$$
Q(v)=c(|v_0|^{-2}v\otimes v_0{}^\dagger_{\rm c}),\quad v\in\Bbb C^2,
\eqno(4.29)
$$
where $c$ has been defined earlier.

The Dirac fermion field $\Psi$ can be thought of as a pair of
Weyl fermion fields $(\Psi^+,\Psi^-)$ with $\Psi^\pm\in
\Pi\Omega^0(X,\Sigma^\pm)$. We set 
$$
\psi^+=\ee^{{3\over 2}\varphi}\tilde{\bar Q}(\Psi^+),
\quad
\psi^-=\ee^{{3\over 2}\varphi}Q(\Psi^-).
\eqno(4.30)
$$
Then, $\psi^\pm\in\Pi(\Omega^0(X,\varpi^\pm)\otimes\Bbb C)$,
where the $\varpi^\pm$ are defined in (2.43).

\vskip.12cm\par\noindent
{\it Proof}. $\SU(2)$  corresponds 
precisely via $c$ to the group $\Sp(1)$ of unit length quaternions in 
$\Bbb H$. Further, as $\det U=1$ and $U=C^{-1}\bar U_{\rm c}C
=U^{-1}{}^\dagger_{\rm c}$ for $U\in\SU(2)$, one has 
$c(U)=\tilde c(U)=\bar c(U)^{-1}$ whenever $U\in\SU(2)$. 
Now, comparing the basic relation
$$
r_{\alpha\beta 0a}1_2+r_{\alpha\beta ea}\tau_e=
\Sigma^+{}_{\alpha\beta}\big(\delta_{0a}1_2+\delta_{ea}\tau_e\big)
(\Sigma^-{}_{\alpha\beta})^{-1},
\eqno(4.31)
$$
satisfied by $\Sigma^\pm_{\alpha\beta}$, and the relation
$$
r_{\alpha\beta 0a}1_2+r_{\alpha\beta ea}\tau_e=
c^{-1}(\theta^+{}_{\alpha\beta})\big(\delta_{0a}1_2+\delta_{ea}\tau_e\big)
c^{-1}((\theta^-{}_{\alpha\beta})^{-1}),
\eqno(4.32)
$$
following from (3.39), and recalling that $c^{-1}(\theta^\pm{}_{\alpha\beta})
\in\SU(2)$ as $\theta^\pm{}_{\alpha\beta}$ is $\Sp(1)$ valued, one concludes 
that
$$
c(\Sigma^\pm{}_{\alpha\beta})=\theta^\pm{}_{\alpha\beta},
\eqno(4.33)
$$
provided the spin structure entering into the definition of 
$\theta^\pm$ is suitably chosen.
Now, from (4.29), one has that $Q(Uv)=c(U)Q(v)$ and $\tilde{\bar Q}(Uv)
=\tilde{\bar Q}(v)c(U)^{-1}$ for $U\in \SU(2)$ and $v\in\Pi\Bbb C^2$.
Hence,
$$
\eqalignno{
\tilde{\bar Q}(\Psi^+{}_\alpha)
&=\tilde{\bar Q}(\Sigma^+{}_{\alpha\beta}\Psi^+{}_\beta)
=\tilde{\bar Q}(\Psi^+{}_\beta)c(\Sigma^+{}_{\beta\alpha})
=\tilde{\bar Q}(\Psi^+{}_\beta)\theta^+{}_{\beta\alpha},
&(4.34)\cr
Q(\Psi^-{}_\alpha)
&=Q(\Sigma^-{}_{\alpha\beta}\Psi^-{}_\beta)
=c(\Sigma^-{}_{\alpha\beta})Q(\Psi^-{}_\beta)
=\theta^-{}_{\alpha\beta}Q(\Psi^-{}_\beta).
&\cr}
$$
From here, it is easy to show the statement combining (4.30) 
and (3.29) and (3.27).
\hfill $QED$ \vskip.12cm

In terms of $\psi^\pm$, the action functional can be written as 
\footnote{}{}
\footnote{${}^6$}{The relative minus sign is due to the anticommuting 
nature of the fields $\Psi^\pm$.}
$$
\eqalignno{
I(\psi^+,\psi^-,\tilde\psi^+,\tilde\psi^-)
&=|v_0|^2{2\over \pi^2}
\real\int_X\Big[\tilde\psi^+\bar\partial_R\wedge\star\hskip 2pt 
dq\hskip 2pt \psi^--\psi^+\bar\partial_R\wedge\star\hskip 2pt 
dq\hskip 2pt \tilde\psi^-\Big]
&(4.35)\cr
&=|v_0|^2{2\over \pi^2}
\real\int_X\Big[\tilde\psi^+\star dq\wedge\bar\partial_L\psi^-
-\psi^+\star dq\wedge\bar\partial_L\tilde\psi^-\Big].
&\cr}
$$

\vskip.12cm\par\noindent
{\it Proof}. Using (3.17) and the formulae 
$$
\gamma_0=i
\Big(\matrix{\hphantom{-}0&1_2\cr-1_2&0\cr}\Big),\quad
\gamma_e=i\Big(\matrix{\hphantom{-}0&\tau_e\cr
-\tau_e{}^\dagger_{\rm c}&0\cr}\Big),
\eqno(4.36)
$$
one can cast the action integral (4.28) as 
$$
\eqalignno{
2\pi^2{\cal I}(\Psi,\Psi^\dagger_{\rm c},e^\vee)
&=+4\real_{\rm c}\int_X\Big[(\ee^{{3\over2}\varphi}
\Psi^+{}^\dagger_{\rm c})
(\partial_01_2+\partial_i\tau_i)_R\ee^{{3\over2}\varphi}\Psi^-\Big]\star 1
&(4.37)\cr
&=-4\real_{\rm c}\int_X\Big[\ee^{{3\over2}\varphi}\Psi^+{}^\dagger_{\rm c}
(\partial_01_2+\partial_i\tau_i)_L(\ee^{{3\over2}\varphi}\Psi^-)\Big]\star 1.
&\cr}
$$
One can show that $\real_{\rm c}(v_2{}^\dagger_{\rm c} v_1)
=|v_0|^2\real\Big(\bar Q(v_2)Q(v_1)-\tilde{\bar Q}(v_2)\tilde Q(v_1)\Big)$ 
for $v_1,~v_2\in\Pi\Bbb C^2$ and that $Q(Uv)=c(U)Q(v)$ and
$\tilde Q(Uv)=c(U)\tilde Q(v)$ for $U\in\SU(2)$ and $v\in\Pi\Bbb C^2$. 
From here, using the relations (2.2), (2.8), (2.9) and (2.44) 
and the definition (4.30), one gets the above result. 
\hfill $QED$ \vskip.12cm

The classical field equations of $\Psi$ are \ref{15}
$$
\gamma_ae_a{}^jD_j\Psi=0.
\eqno(4.38)
$$
In terms of $\psi^\pm$, they read simply as 
$$
\psi^+\bar\partial_R=0,\quad \bar\partial_L\psi^-=0.
\eqno(4.39)
$$
Hence, $\psi^+$ ($\psi^-$) is right (left) Fueter holomorphic.
See the discussion of section 3 concerning the solutions of these
equations. 

The energy--momentum tensor of the Dirac fermion $\Psi$ is \ref{15}
$$
\eqalignno{
{\cal T}_a(\Psi,\Psi^\dagger_{\rm c},e^\vee)
&=\real_{\rm c}\Big\{{1\over 2i}\Psi^\dagger_{\rm c}\Big[
\gamma_be^\vee{}_{bj}e_a{}^kD_k+\gamma_aD_j
&(4.40)\cr
&\hphantom{=}-{1\over 2}e^\vee{}_{aj}\gamma_be_b{}^kD_k
+{1\over 2}[\gamma_a,\gamma_c]
e^\vee{}_{cj}\gamma_be_b{}^kD_k\Big]\Psi\Big\}dx^j.
&\cr}
$$
It is straightforward though a bit lengthy to check that 
(4.8) holds. The conformally invariant 
 energy--momentum tensor $T$ can be computed. One finds 
$$
\eqalignno{
T(\psi^+,\psi^-,\tilde\psi^+,\tilde\psi^-)
&={|v_0|^2\over 4}\Big\{
-\tilde\psi_-d\psi_++\psi_-d\tilde\psi_+ 
+d\tilde\psi_-\psi_+-d\psi_-\tilde\psi_+
&(4.41)\cr
&\hphantom{=}
+d\bar q\big(\tilde{\bar\psi}_+\bar\psi_- 
-\bar\psi_+\tilde{\bar\psi}_-\big)\partial_{qR}
+\big(\psi_-\tilde\psi_+-\tilde\psi_-\psi_+\big)\partial_{\bar qR}d\bar q
\vphantom{3\over 2}
&\cr
&\hphantom{=}
+{3\over 2}\Big(\partial_{qL}\bar\psi_+\tilde{\bar\psi}_- 
-\partial_{qL}\tilde{\bar\psi}_+\bar\psi_-
+\tilde\psi_-\psi_+\partial_{\bar qR}
-\psi_-\tilde\psi_+\partial_{\bar qR}\Big)d\bar q
&\cr
&\hphantom{=}
+{1\over 2}d\bar q\Big(\bar\psi_+\tilde{\bar\psi}_-\partial_{qR} 
-\tilde{\bar\psi}_+\bar\psi_-\partial_{qR}
+\partial_{\bar qL}\tilde\psi_-\psi_+
-\partial_{\bar qL}\psi_-\tilde\psi_+\Big)
\Big\}
&\cr
&={|v_0|^2\over 4}\Big\{
-\tilde\psi_-d\psi_++\psi_-d\tilde\psi_+ 
+d\tilde\psi_-\psi_+-d\psi_-\tilde\psi_+
&\cr
&\hphantom{=}
-\partial_{qL}\big(\tilde{\bar\psi}_+\bar\psi_- 
-\bar\psi_+\tilde{\bar\psi}_-\big)d\bar q
-d\bar q\partial_{\bar qL}\big(\psi_-\tilde\psi_+-\tilde\psi_-\psi_+\big)
\vphantom{3\over 2}
&\cr
&\hphantom{=}
-{1\over 2}\Big(\partial_{qL}\bar\psi_+\tilde{\bar\psi}_- 
-\partial_{qL}\tilde{\bar\psi}_+\bar\psi_-
+\tilde\psi_-\psi_+\partial_{\bar qR}
-\psi_-\tilde\psi_+\partial_{\bar qR}\Big)d\bar q
&\cr
&\hphantom{=}
-{3\over 2}d\bar q\Big(\bar\psi_+\tilde{\bar\psi}_-\partial_{qR} 
-\tilde{\bar\psi}_+\bar\psi_-\partial_{qR}
+\partial_{\bar qL}\tilde\psi_-\psi_+
-\partial_{\bar qL}\psi_-\tilde\psi_+\Big)
\Big\}.
&\cr}
$$
We have checked that $T$ fulfills (4.10) and that (4.11) and (4.12) 
hold, when the $\psi^\pm$ fulfills the field equations (4.39).
For a field configuration $\psi^\pm$ satisfying (4.39),
the terms proportional to $d\bar q$
vanish identically, simplifying the above expressions.

The Dirac action has an obvious ${\rm U}(1)$ symmetry. The corresponding 
${\rm U}(1)$ current is
$$
{\cal J}(\Psi,\Psi^\dagger_{\rm c},e^\vee)
=\Psi^\dagger_{\rm c}\gamma_ae^\vee{}_{aj}\Psi dx^j.
\eqno(4.42)
$$
It is easy to see that (4.17) is fulfilled with
$$
J(\psi^+,\psi^-,\tilde\psi^+,\tilde\psi^-)
=|v_0|^2\real\Big(i\tilde\psi_+dq\psi_-
+i\psi_+dq\tilde\psi_-\Big).
\eqno(4.43)
$$
$J$ satisfies (4.18), when the $\psi^\pm$ satisfies the field equations 
(4.39).
\par\vskip.6cm
\item{\bf 5.} {\bf Quantum 4--dimensional conformal
field theory and Kulkarni geometry}
\vskip.4cm
\par
In this section, we consider first some general properties
of a conformal quantum field theory on a Kulkarni 4--fold $X$
concentrating on the quantum energy--momentum tensor. 
We then analyze the properties of the operator product expansions for 
the simple free models studied in the previous section.

Below, we shall assume that $X$ is compact.

In a quantum 4--dimensional conformal field theory, the local classical 
action ${\cal I}(\Phi,e^\vee)$ is affected by quantum corrections. The 
resulting effective action ${\cal I}_{\rm e}(\Phi,e^\vee)$ is a non local
functional of $\Phi$ and $e^\vee{}_a$. In general, 
${\cal I}_{\rm e}(\Phi,e^\vee)$
is no longer conformally invariant but, conversely,  suffers an additive
conformal anomaly. We assume that, for any smooth function $f$, 
$$
{\cal I}_{\rm e}(\ee^{-f\Lambda}\Phi,\ee^fe^\vee)
={\cal I}_{\rm R}(f,e^\vee)+{\cal I}_{\rm e}(\Phi,e^\vee),
\eqno(5.1)
$$
where ${\cal I}_{\rm R}(f,e^\vee)$ is the Riegert action, which is 
local and independent from $\Phi$ \ref{16--17}.

{\it The quantum energy--momentum tensor}

One can define the energy--momentum 
tensors ${\cal T}_{{\rm e}a}(\Phi,e^\vee)$ 
and ${\cal T}_{{\rm R}a}(f,e^\vee)$ for the actions ${\cal I}_{\rm e}$ and 
${\cal I}_{\rm R}$ in the same way as done in the classical case:
$\delta_{e^\vee}{\cal I}_{\rm e}
=-{1\over 2\pi^2}\int_X\langle {\cal T}_{{\rm e}a},\delta e_a\rangle *1$ 
and $\delta_{e^\vee}{\cal I}_{\rm R}
=-{1\over 2\pi^2}\int_X\langle {\cal T}_{{\rm R}a},\delta e_a\rangle *1$,
where $\delta_{e^\vee}f=-{1\over 4}\delta\ln e$.
Because of the conformal anomaly, ${\cal T}_{{\rm e}a}$ and 
${\cal T}_{{\rm R}a}$ do not satisfy a condition of tracelessness
analogous to (4.4). However, since invariance under the automorphism group 
of the orthonormal frame bundle is not anomalous, ${\cal T}_{{\rm e}a}$ 
still satisfies (4.5) and (4.6) in the vacuum, i. e. at vanishing field 
configurations. So, ${\cal T}_{{\rm e}a}|_{\Phi=0}$ is symmetric, 
$$
{\cal T}_{{\rm e}a}\wedge e^\vee{}_a|_{\Phi=0}=0,
\eqno(5.2)
$$
and satisfies the Ward identity 
$$
\big(d*{\cal T}_{{\rm e}a}
+\omega_{ab}\wedge *{\cal T}_{{\rm e}b})|_{\Phi=0}=0.
\eqno(5.3)
$$
${\cal T}_{{\rm R}a}$ is also symmetric 
$$
{\cal T}_{{\rm R}a}\wedge e^\vee{}_a=0,
\eqno(5.4)
$$
while its Ward identity reads
$$
d*{\cal T}_{{\rm R}a}+\omega_{ab}\wedge *{\cal T}_{{\rm R}b}
+\Big({1\over 4}d{\cal C}_{\rm R}-{\cal C}_{\rm R}df\Big)\wedge *e^\vee{}_a
=0,
\eqno(5.5)
$$
where the functional ${\cal C}_{\rm R}(f,e^\vee)$ is defined by
$\delta_f{\cal I}_{\rm R}=+{1\over 2\pi^2}\int_X{\cal C}_{\rm R}\delta f*1$
The origin of the extra terms in the Ward identity 
(5.5) is easily understood. If ${\cal I}_{\rm R}$ were the classical 
action of some automorphism invariant field theory,  they 
would be absent for a field $f$ 
satisfying the classical field equation ${\cal C}_{\rm R}=0$
and (5.5) would be analogous to (4.6). 

There is another piece of information that is relevant and does not
follow directly from (5.1). One has 
$$
{\cal C}_{\rm R}(0,e^\vee)=0\quad\hbox{on any open set of $X$ 
where $R_{ab}=0$}.
\eqno(5.6)
$$
This identity can be justified by noting that, on dimensional grounds,
${\cal C}_{\rm R}(0,e^\vee)$ is the sum of two contributions. 
The first is quadratic in the components of the Riemann 2--form
$R_{ab}$
and the derived forms. The second is proportional to $d*ds$, where
$s$ is the Ricci scalar. Both contributions vanish in the regions 
where the background $e^\vee{}_a$ is flat.

Because of the anomalous breaking of conformal invariance in
the quantum theory, ${\cal T}_{{\rm e}a}$ does not satisfies a relation 
of the form (4.7) in the locally conformally flat background of eq. (3.2)
and therefore it does not have a structure like that 
exhibited in (4.8). However, it is still possible extract from 
${\cal }T_{{\rm e}a}$ a part $T_{{\rm e}i}(\phi)$ 
depending only on $\phi$ and 
the conformal geometry of the base manifold $X$. Indeed,
$$
{\cal T}_{{\rm e}a}(\Phi,e^\vee)
=\ee^{-3\varphi}\Big[\delta_{ai}T_{{\rm e}i}(\phi)
+{\cal L}_{{\rm R}a}(\varphi,\ee^{-\varphi}e^\vee)\Big],
\eqno(5.7)
$$
where 
$$
{\cal L}_{{\rm R}a}={\cal T}_{{\rm R}a}+{1\over 4}{\cal C}_{\rm R}e^\vee{}_a.
\eqno(5.8)
$$
Since the action ${\cal I}_{\rm R}$ is local, 
${\cal T}_{{\rm R}a}$ and ${\cal C}_{\rm R}$ are local expression in
the fields $f$ and $e^\vee{}_a$ involving no integration on $X$. 
They are therefore defined also when $f$ and $e^\vee{}_a$ are replaced by the 
local scale $\varphi$ and the local dual vierbein
$\ee^{-\varphi}e^\vee{}_a$. The covariance of the composite fields 
obtained in this way is however quite different from the original one, 
as will be shown in a moment. 
Now, one can verify that $T_{{\rm e}i}(\phi)$ is conformally invariant,
as suggested by the notation. Following (4.9), one sets
$$
T_{\rm e}={1\over 4}(T_{{\rm e}0}-T_{{\rm e}r}j_r).
\eqno(5.9)
$$
Then, one can verify that $T_{\rm e}$ is traceless:
$$
\real\big(T_{\rm e}\iota(\partial_{\bar q})\big)=0.
\eqno(5.10)
$$
Further, in the vacuum, i. e. when $\phi=0$, $T_{\rm e}$ is symmetric
and conserved, so that 
$$
\real\big(dq\wedge T_{\rm e})|_{\phi=0}=0
\eqno(5.11)
$$
and
$$
d\star T_{\rm e}|_{\phi=0}=0.
\eqno(5.12)
$$

\vskip.12cm\par\noindent
{\it Proof}. We give only a sketch of the proof. 
By varying (5.1) with respect to $f$ and $e^\vee{}_a$, one obtains 
$$
\ee^{3f}\langle{\cal T}_{{\rm e}a}(\ee^{-f\Lambda}\Phi,\ee^fe^\vee)
,e_a\rangle-{\cal C}_{\rm R}(f,e^\vee)=0,
\eqno(5.13)
$$
$$
\ee^{3f}{\cal T}_{{\rm e}a}(\ee^{-f\Lambda}\Phi,\ee^fe^\vee)
-{\cal T}_{{\rm e}a}(\Phi,e^\vee)
-{\cal T}_{{\rm R}a}(f,e^\vee)
-{1\over 4}{\cal C}_{\rm R}(f,e^\vee)e^\vee{}_a=0.
\eqno(5.14)
$$
From (5.1), it follows that the action ${\cal I}_{\rm R}$
satisfies the so called 1--cocycle relation 
$$
{\cal I}_{\rm R}(f_1+f_2,e^\vee)
-{\cal I}_{\rm R}(f_1,\ee^{f_2}e^\vee)-{\cal I}_{\rm R}(f_2,e^\vee)=0,
\eqno(5.15)
$$
for any two smooth functions $f_1$, $f_2$. 
By varying this identity with respect to $f_1$, $f_2$ and $e^\vee{}_a$, one
obtains 
$$
{\cal C}_{\rm R}(f_1+f_2,e^\vee)
-\ee^{4f_2}{\cal C}_{\rm R}(f_1,\ee^{f_2}e^\vee)=0,
\vphantom{1\over 4}
\eqno(5.16)
$$
$$
\ee^{3f_2}\langle{\cal T}_{{\rm R}a}(f_1,\ee^{f_2}e^\vee),e_a\rangle
+{\cal C}_{\rm R}(f_2,e^\vee)=0,
\eqno(5.17)
$$
$$
{\cal T}_{{\rm R}a}(f_1+f_2,e^\vee)
-\ee^{3f_2}{\cal T}_{{\rm R}a}(f_1,\ee^{f_2}e^\vee)
-{\cal T}_{{\rm R}a}(f_2,e^\vee)
-{1\over 4}{\cal C}_{\rm R}(f_2,e^\vee)e^\vee{}_a=0.
\eqno(5.18)
$$
Define ${\cal E}_a(\phi,\varphi,e^\vee)=
\ee^{3\varphi}{\cal T}_{{\rm e}a}(\ee^{-\varphi\Lambda}\phi,e^\vee)
-{\cal L}_{{\rm R}a}(\varphi,\ee^{-\varphi}e^\vee)$. Using (5.14) with 
$\Phi$ substituted by $\ee^{-\varphi\Lambda}\phi$ and (5.16) and (5.18) 
with $f_1$, $f_2$ and $e^\vee{}_a$ substituted by $f$, $\varphi$ and 
$\ee^{-\varphi}e^\vee{}_a$, respectively, one verifies that
${\cal E}_a(\phi,\varphi+f,\ee^fe^\vee)
={\cal E}_a(\phi,\varphi,e^\vee)$,
showing the conformal invariance of ${\cal E}_a(\phi,\varphi,e^\vee)$. 
Thus, 
$$
T_{{\rm e}i}(\phi)=\delta_{ia}\Big[
\ee^{3\varphi}{\cal T}_{{\rm e}a}(\ee^{-\varphi\Lambda}\phi,e^\vee)
-{\cal L}_{{\rm R}a}(\varphi,\ee^{-\varphi}e^\vee)\Big]
\eqno(5.19)
$$
depends only on $\phi$ and the background conformal geometry.
Using (5.13) with $\Phi$, $f$ and $e^\vee{}_a$ replaced by
$\phi$, $\varphi$ and $\ee^{-\varphi}e^\vee{}_a$ and (5.17) with
$f_1$, $f_2$ and $e^\vee{}_a$ substituted by $\varphi$, $0$  and 
$\ee^{-\varphi}e^\vee{}_a$, respectively, one verifies that
$\delta_{ia}\iota(\ee^{\varphi}e_a)T_{{\rm e}i}(\phi)=
{\cal C}_{\rm R}(0,\ee^{-\varphi}e^\vee)$. 
${\cal C}_{\rm R}(0,\ee^{-\varphi}e^\vee)=0$, by (5.6),
because, by (3.2), the local background $\ee^{-\varphi}e^\vee{}_a$ is flat.
So, $\delta_{ia}\iota(\ee^{\varphi}e_a)T_{{\rm e}i}(\phi)=0$. 
This relation yields (5.10) immediately upon using (5.9) and recalling (3.1) 
and (3.5). Finally, from  (5.2) and (5.4), we obtain the symmetry 
relation $\delta_{ia}T_{{\rm e}i}(0)\wedge\ee^{-\varphi}e^\vee{}_a=0$.
From here, (5.11) follows upon using (5.9) and recalling (3.2) and (3.7). 
Next, by using the symmetry relation (5.2) and the Ward identity (5.3)
and exploiting  relations (3.9) and (3.17), one has 
$$
\eqalignno{
d\star\Big[\ee^{3\varphi}{\cal T}_{{\rm e}a}(0,e^\vee)\Big]
&=\ee^\varphi\Big[d\varphi\wedge *{\cal T}_{{\rm e}a}(0,e^\vee)+
d*{\cal T}_{{\rm e}a}(0,e^\vee)\Big]
&(5.20)\cr
&=\ee^\varphi\langle {\cal T}_{{\rm e}b}(0,e^\vee),e_b\rangle
d\varphi\wedge *e^\vee{}_a.\vphantom{1\over 2}
&\cr}
$$
From the Ward identity (5.5) with $f$ and $e^\vee{}_a$ replaced by 
$\varphi$ and $\ee^{-\varphi}e^\vee{}_a$, one deduces further that
$$
d\star{\cal L}_{{\rm R}a}(\varphi,\ee^{-\varphi}e^\vee)
={\cal C}_{\rm R}(\varphi,\ee^{-\varphi}e^\vee)d\varphi
\wedge\star\ee^{-\varphi}e^\vee{}_a.
\eqno(5.21)
$$
In deriving this relation, one uses that $d\star(\ee^{-\varphi}e^\vee{}_a)
=0$, by (3.2). Now, by (5.19), $d\star T_{{\rm e}i}(0)$ is given by 
the difference of the left hand sides of eqs. (5.20) and (5.21), which  
vanishes by (5.13) with $\Phi$, $f$ and $e^\vee{}_a$ replaced by $0$, 
$\varphi$ and $\ee^{-\varphi}e^\vee{}_a$ and by (3.2) and (3.9). Hence,
$d\star T_{{\rm e}i}(0)=0$. From here, using (5.9), (5.12) follows. 
\hfill $QED$ \vskip.12cm

\par\noindent
The above treatment is essentially a reformulation of the classic 
results of ref. \ref{18} highlighting the connection with Kulkarni
geometry.

As noticed earlier, $T_{\rm e}$ does not transform as its classical 
counterpart under coordinate changes. In fact, on 
$U_\alpha\cap U_\beta\not=\emptyset$, one has
$$
T_{{\rm e}\alpha}=\zeta_{3\alpha\beta}
\big(T_{{\rm e}\beta}+\varrho_{\alpha\beta}\big),
\eqno(5.22)
$$
where $\zeta_3$ is defined in (2.34) and 
$$
\varrho_{\alpha\beta}
={1\over 4}\Big[{\cal L}_{{\rm R}\beta 0}(dx_\beta,
\ln(|\eta^+{}_{\alpha\beta}|/|\eta^-{}_{\alpha\beta}|))
-{\cal L}_{{\rm R}\beta e}(dx_\beta,
\ln(|\eta^+{}_{\alpha\beta}|/|\eta^-{}_{\alpha\beta}|))
j_e\Big].
\eqno(5.23)
$$

\vskip.12cm\par\noindent
{\it Proof}. Set $t_{\alpha\beta}
=\ln(|\eta^+{}_{\alpha\beta}|/|\eta^-{}_{\alpha\beta}|)$. Then, 
$$
\eqalignno{
{\cal T}_{{\rm R}\alpha a}(\varphi_\alpha,
\ee^{-\varphi_\alpha}e^\vee{}_\alpha)
&=\ee^{3\varphi_\alpha}\Big[
{\cal T}_{{\rm R}\alpha a}(0,e^\vee{}_\alpha)
-{\cal T}_{{\rm R}\alpha a}(-\varphi_\alpha,e^\vee{}_\alpha)
\vphantom{1\over 2}&(5.24)\cr
&\hphantom{=}
-{1\over 4}{\cal C}_{\rm R}(-\varphi_\alpha,e^\vee{}_\alpha)
e^\vee{}_{\alpha a}\Big]
\vphantom{1\over 2}&\cr
&=\ee^{3\varphi_\beta-3t_{\alpha\beta}}r_{\alpha\beta ab}\Big[
{\cal T}_{{\rm R}\beta b}(0,e^\vee{}_\beta)
-{\cal T}_{{\rm R}\beta b}(-\varphi_\beta+t_{\alpha\beta},e^\vee{}_\beta)
\vphantom{1\over 2}&\cr
&\hphantom{=}
-{1\over 4}{\cal C}_{\rm R}(-\varphi_\beta+t_{\alpha\beta},e^\vee{}_\beta)
e^\vee{}_{\beta b}\Big]
\vphantom{1\over 2}&\cr
&=\ee^{3\varphi_\beta-3t_{\alpha\beta}}r_{\alpha\beta ab}\Big[
\ee^{-3\varphi_\beta}
{\cal T}_{{\rm R}\beta b}(\varphi_\beta,\ee^{-\varphi_\beta}e^\vee{}_\beta)
+{\cal T}_{{\rm R}\beta b}(-\varphi_\beta,e^\vee{}_\beta)
\vphantom{1\over 2}&\cr
&\hphantom{=}
+{1\over 4}{\cal C}_{\rm R}(-\varphi_\beta,e^\vee{}_\beta)e^\vee{}_{\beta b}
-{\cal T}_{{\rm R}\beta b}(-\varphi_\beta+t_{\alpha\beta},e^\vee{}_\beta)
\vphantom{1\over 2}&\cr
&\hphantom{=}
-{1\over 4}{\cal C}_{\rm R}(-\varphi_\beta+t_{\alpha\beta},e^\vee{}_\beta)
e^\vee{}_{\beta b}\Big]
\vphantom{1\over 2}&\cr
&=\ee^{-3t_{\alpha\beta}}r_{\alpha\beta ab}\Big[
{\cal T}_{{\rm R}\beta b}(\varphi_\beta,\ee^{-\varphi_\beta}e^\vee{}_\beta)
-{\cal T}_{{\rm R}\beta b}(t_{\alpha\beta},\ee^{-\varphi_\beta}e^\vee{}_\beta)
\vphantom{1\over 2}&\cr
&\hphantom{=}
-{1\over 4}{\cal C}_{\rm R}(t_{\alpha\beta},\ee^{-\varphi_\beta}e^\vee{}_\beta)
\ee^{-\varphi_\beta}e^\vee{}_{\beta b}\Big],
\vphantom{1\over 2}&\cr}
$$
where $r_{\alpha\beta}$ is the same $\SO(4)$ valued function as that
appearing in (3.38).
Here, the first identity is proven by applying (5.18) with $f_1$, $f_2$
and $e^\vee{}_a$ substituted by $\varphi_\alpha$, $-\varphi_\alpha$
and $e^\vee{}_{\alpha a}$, respectively. The second identity follows from 
(3.29), (3.38) and the relation 
$$
{\cal T}_{{\rm R}\alpha a}=r_{\alpha\beta ab}{\cal T}_{{\rm R}\beta b}
\eqno(5.25)
$$
analogous to (4.14). The third identity is proven by applying (5.18) 
with $f_1$, $f_2$ and $e^\vee{}_a$ substituted by $\varphi_\beta$, 
$-\varphi_\beta$ and $e^\vee{}_{\beta a}$, respectively. The fourth and final
identity is shown by applying (5.17) and (5.18) with $f_1$, $f_2$ and 
$e^\vee{}_a$ substituted by $t_{\alpha\beta}$, $-\varphi_\beta$ and 
$e^\vee{}_{\beta a}$, respectively. Next, one has 
$$
\eqalignno{
{\cal C}_{\rm R}(\varphi_\alpha,\ee^{-\varphi_\alpha}e^\vee{}_\alpha)
\ee^{-\varphi_\alpha}e^\vee{}_{\alpha a}
&=\ee^{3\varphi_\alpha}{\cal C}_{\rm R}(0,e^\vee{}_\alpha)
e^\vee{}_{\alpha a}
\vphantom{1\over 4}&(5.26)\cr
&=\ee^{3\varphi_\beta-3t_{\alpha\beta}}r_{\alpha\beta ab}
{\cal C}_{\rm R}(0,e^\vee{}_\beta)e^\vee{}_{\beta b}
\vphantom{1\over 4}&\cr
&=\ee^{-3t_{\alpha\beta}}r_{\alpha\beta ab}
{\cal C}_{\rm R}(\varphi_\beta,\ee^{-\varphi_\beta}e^\vee{}_\beta)
\ee^{-\varphi_\beta}e^\vee{}_{\beta b}.
\vphantom{1\over 4}&\cr}
$$
The first identity is obtained by applying (5.18) with $f_1$, $f_2$
and $e^\vee{}_a$ substituted by $\varphi_\alpha$, $-\varphi_\alpha$
and $e^\vee{}_{\alpha a}$, respectively. The second identity follows from 
(3.29) and (3.38). The third identity is proven by applying (5.17) 
with $f_1$, $f_2$ and $e^\vee{}_a$ substituted by $\varphi_\beta$, 
$-\varphi_\beta$ and $e^\vee{}_{\beta a}$, respectively.
Combining (3.27), (3.29), (3.39), (5.24) and (5.26) with (5.8) and (5.19)
and recalling (2.34), 
one checks easily that the matching relation of the $T_{{\rm R}\alpha}$ 
is given by (5.22)--(5.23). \hfill $QED$ \vskip.12cm

\par\noindent
The compatibility of (5.22) and (5.10)--(5.12) entails the following 
relations
$$
\real\big(\varrho_{\alpha\beta}\iota(\partial_{\bar q\beta})\big)=0,
\eqno(5.27)
$$
$$
\real(dq_\beta\wedge\varrho_{\alpha\beta})=0,
\eqno(5.28)
$$
$$
d\star_\beta\varrho_{\alpha\beta}=0.
\eqno(5.29)
$$

\vskip.12cm\par\noindent
{\it Proof}. The verification of () and () is completely straightforward.
To show (), one has take into account the fact that, if local
quaternionic 1--forms $\nu_\alpha$ satisfy 
$\real\big(\nu_\alpha\iota(\partial_{\bar q\alpha})\big)=0$ and 
$\real(dq_\alpha\wedge\nu_\alpha)=0$, then the equation 
$d\star_\alpha\nu_\alpha=0$ is covariant under the matching relation
$\nu_\alpha=\zeta_{3\alpha\beta}\nu_\beta$. 
\hfill $QED$ \vskip.12cm

From (5.23), it appears that $\varrho_{\alpha\beta}$ depends only on the 
underlying conformal geometry. So, the matching relation (5.22) is completely
analogous to that of the conformally invariant energy--momentum tensor 
in 2--dimensional conformal field theory and $\varrho_{\alpha\beta}$
is a 4--dimensional generalization of the Schwarzian derivative.

The form of the conformal anomaly \ref{19--20}
is determined up to a term 
of the form $\delta{\cal K}(e^\vee)$, where $\delta$ denotes variation 
with respect to the scale of $e^\vee{}_a$ and ${\cal K}(e^\vee)$ is a local 
functional of $e^\vee{}_a$. The form of the anomaly can be rendered
simpler by means of a convenient choice of ${\cal K}$. A further 
simplification is yielded by the local conformal flatness of 
the background $e^\vee{}_a$ of eq. (3.2), 
which makes the contribution containing 
the square of the Weyl tensor vanish identically. In this way the 
conformal anomaly can be cast as
$$
\delta{\cal I}_{\rm e}={\kappa\over 128\pi^2}\int_X
\Big[32\pi^2\epsilon-{2\over 3}d*ds\Big]\langle\delta e^\vee{}_a,e_a\rangle,
\eqno(5.30)
$$
where $\epsilon$ is the Euler density, defined above (3.25), 
and $s$ is the Ricci scalar.
$\kappa$ is a real coefficient called central charge. In fact, the 
expression of the anomaly is simpler than it looks at first glance. A 
detailed calculation, exploiting the local conformal flatness 
of $e^\vee{}_a$, shows that it can be written in the form
$$
\delta{\cal I}_{\rm e}
={32\kappa\over\pi^2}\int_X\square\star\square\varphi\delta\varphi,
\eqno(5.31)
$$
where $\square={1\over 16}d\star d$ is the D'Alembert operator.
In this form, the similarity with the standard 2--dimensional case
is apparent. As a byproduct, we learn also that 
$\square\star\square\varphi$ belongs to $\Omega^4(X)$, 
an interesting geometric result.

The Riegert action corresponding to the anomaly (5.19) is given by 
\ref{16--17}
$$
\eqalignno{
{\cal I}_{\rm R}(f,e^\vee)
&={\kappa\over 16\pi^2}\int_X\Big[d*df\wedge *d*df
-{2\over 3}sdf\wedge *df
&(5.32)\cr
&\hphantom{=}+2e_a(f)S_a\wedge *df
+\Big(16\pi^2\epsilon-{1\over 3}d*ds\Big)f\Big].
&\cr}
$$
In the locally conformally flat background $e^\vee{}_a$ of eq. (3.2),
${\cal I}_{\rm R}$ can be written as 
$$
{\cal I}_{\rm R}(f,\varphi)
={32\kappa\over\pi^2}\int_X
\Big[{1\over 2}f\square\star\square f
+\square\star\square\varphi f\Big].
\eqno(5.33)
$$
When written in this form, the resemblance of the 4--dimensional
Riegert action and the well--known 2--dimensional Liouville action is
striking. The calculation shows also that $\square\star\square$
is a globally defined differential operator of order 4
mapping $\Omega^0(X)$ into $\Omega^4(X)$ \footnote{}{}
\footnote{${}^7$}{This operator, as many others,
could have been included in the list of the 
natural differential operators of a Kulkarni 4--fold studied in section 2.
To keep the size of this paper reasonable, we decided to 
limit our discussion to $\square$ and $\bar\partial_{R,L}$.}.

It is now straightforward though quite tedious to compute $T_{\rm e}$.
Set
$$
\eqalignno{
P(f)&=4df\partial_q\star\square f
+4\partial_qfd\star\square f
-{1\over 12}d\partial_q\star(df\wedge\star df)
&(5.34)\cr
&\hphantom{=}
+{1\over 2}dx^i\star(d\partial_{xi}f\wedge
\star d\partial_qf)-8d\partial_qf\star\square f
-{4\over 3}d\partial_q\star\square f
&\cr
&\hphantom{=}
+d\bar q\Big[8(\star\square f)^2-
{1\over 6}\star\square\star(df\wedge\star df)
+{16\over 3}\star\square\star\square f\Big].
&\cr}
$$
Then, $T_{\rm e}(\phi)$ is given by
$$
T_{\rm e}(\phi)=\ee^{3\varphi}
{\cal T}_{\rm e}(\ee^{-\varphi\Lambda}\phi,e^\vee)
-\kappa P(\varphi),
\eqno(5.35)
$$
where ${\cal T}_{\rm e}={1\over 4}({\cal T}_{{\rm e}0}
-{\cal T}_{{\rm e}e}j_e)$. 
The 4--dimensional Schwarzian derivative $\varrho_{\alpha\beta}$
defined in (5.23) is given explicitly by
$$
\varrho_{\alpha\beta}
=\kappa P_\beta(\ln(|\eta^+{}_{\alpha\beta}|/|\eta^-{}_{\alpha\beta}|)).
\eqno(5.36)
$$

{\it The operator product expansions}

We shall now analyze the structure of the operator product expansions
for the simple free models studied in section 4. 

Consider the complex boson $\Phi$ described by the action (4.19).
The quantum theory is best formulated in terms of the conformally
invariant field $\phi$ governed by the action (4.21).
Inside normalized conformally invariant quantum corelators, the 
classical field equations (4.23) hold up to contact terms
$$
\square\phi=0\quad\hbox{up to contact terms}.
\eqno(5.37)
$$
Hence, the corelators are harmonic in the insertion points of the field 
$\phi$ and its complex conjugate, provided such points remains distinct. 
Since a real harmonic function can be expressed as the real part 
of a Fueter holomorphic function \ref{7}, Fueter analyticity is relevant 
in this model. 
From the form of the action (4.21), it follows in particular that
$$
\eqalignno{
-{8\over\pi^2}\phi(q_2)\square_1\bar\phi_{\rm c}(q_1)
&=\delta^4(q_2-q_1)\star 1_2,
&(5.38)\cr 
-{8\over\pi^2}\square_2\phi(q_2)\bar\phi_{\rm c}(q_1)
&=\delta^4(q_2-q_1)\star 1_1.
&\cr}
$$
This relation can be easily integrated on a given coordinate patch,
yielding
$$
\phi(q_2)\bar\phi_{\rm c}(q_1)={1\over 2|q_2-q_1|^2}+
\hbox{regular harmonic terms}.
\eqno(5.39)
$$

\vskip.12cm\par\noindent
{\it Proof}. From distribution theory, one can show easily that 
$\partial_{\bar q}\partial_q|q-q_0|^{-2}=-{\pi^2\over 4}\delta^4(q-q_0)$
in ${\cal D}'(\Bbb H)$. Further, it is known \ref{7} that there is no
singular harmonic function less singular than $|q-q_0|^{-2}$.
\hfill $QED$ \vskip.12cm

Consider the Dirac fermion $\Psi$ described by the action (4.28).
It is more convenient to formulate the quantum theory 
in terms of the conformally invariant fields $\psi^\pm$ 
governed by the action (4.35).
(We assume $|v_0|=1$ here for the sake of simplicity).
Inside normalized conformally invariant quantum corelators, the 
classical field equations (4.39) hold up to contact terms
$$
\psi^+\bar\partial_R=0,\quad \bar\partial_L\psi^-=0
\quad\hbox{up to contact terms}.
\eqno(4.39)
$$
Hence, the quantum corelators are right (left) Fueter holomorphic in 
the insertion points of the field $\psi^+$ ($\psi^-$), provided
such points do not coincide. This statement must carry a warning. 
Since the fields $\psi^\pm$ and the Fueter operators $\bar\partial_{R,L}$ 
are valued in the non commutative quaternion field, the statement holds 
provided $\bar\partial_R$ ($\bar\partial_L$) acts on $\psi^+$ ($\psi^-$)
within the corelators. The above shows the relevance of Fueter 
analyticity in the present fermionic model.
From the form of the action (4.35), it follows in particular that 
$$
\eqalignno{
{2\over\pi^2}\psi^-(q_2)\tilde\psi^+(q_1)\bar\partial_{R1}
&=\delta^4(q_2-q_1)d\bar q_2,
&(5.41)\cr
{2\over\pi^2}\bar\partial_{L2}\psi^-(q_2)\tilde\psi^+(q_1)
&=\delta^4(q_2-q_1)d\bar q_1.
&\cr}
$$
This relation can be integrated on any given coordinate patch,
producing
$$
\psi^-(q_2)\tilde\psi^+(q_1)=
{\bar q_2-\bar q_1\over |q_2-q_1|^4}
+\hbox{terms right (left) Fueter holomorphic in $q_1$ ($q_2$)}.
\eqno(5.42)
$$

\vskip.12cm\par\noindent
{\it Proof}. From distribution theory, one knows that
$[(\bar q-\bar q_0)|q-q_0|^{-4}]\partial_{\bar qR}
=\partial_{\bar qL}[(\bar q-\bar q_0)|q-q_0|^{-4}]
={\pi^2\over 2}\delta^4(q-q_0)$ in ${\cal D}'(\Bbb H)$. 
Further, it is known \ref{7}
that there is no singular left/right Fueter analytic function that is less
singular than $(\bar q-\bar q_0)|q-q_0|^{-4}$.
\hfill $QED$ \vskip.12cm

The above analysis shows that Fueter analyticity provides useful information 
on the structure of the operator product expansions of free fields. 
It remains to be seen if this will be of any help in computations. 
\par\vskip.6cm
\item{\bf 6.} {\bf Conclusions and outlook}
\vskip.4cm
\par
In the first part of 
this paper, we have tried to formulate the theory of Kulkarni 4--folds in 
a way that parallels as much as possible the customary formulation
of the theory of Riemann surfaces, highlighting in this way their analogies.
This has been possible thanks to the existence of an integrable quaternionic 
structure and of an associated natural notion of analyticity, Fueter 
analyticity. We have also seen that a Kulkarni 4--folds is
equipped with a canonical conformal equivalence class of locally 
conformally flat metrics and that the Riemannian geometry of such metrics 
is particularly simple. 

In the second part of the paper, we have argued that Kulkarni geometry
is the natural geometry of 4--dimensional conformal field theory by showing 
that the action functional, the field equations, the energy--momentum tensor 
and its Ward identity and the operator product expansions take a simple form 
for a conformal field theory on a Kulkarni 4--fold. 

We have not analyzed yet the implications of the geometric setting on the 
operator product expansion of the energy--momentum tensor. This matter is
left for future work \ref{21}. We believe in fact that the customary 
energy--momentum tensor, describing the response of the system to an 
arbitrary variation of an arbitrary background metric, might 
not be the relevant geometric field. One should consider instead 
a modified energy--momentum tensor representing the response of the system 
to an arbitrary variation of an arbitrary locally conformally flat background
metric preserving local conformal flatness. This would be the true analogue 
of the energy--momentum tensor of 2--dimensional conformal field theory, as 
for a 4--fold admitting locally conformally flat metrics, unlike for 
a 2--fold, not all metrics are automatically locally conformally flat.
One may speculate that the improved energy--momentum tensor just described
might obey operator product expansion of universal form as in 2 dimensions.
This remains to be seen. In any case, to carry out the above project requires
the elaboration of the Kulkarni analogue of the Beltrami parametrization 
of conformal structures, a major mathematical task in itself with 
ramifications also in geometry. 
\vfill\eject
\vskip.6cm
\par\noindent
{\bf Acknowledgements.} We are indebt to R. Stora, R. Balbinot
F. Bastianelli and F. Ravanini for helpful discussions.
\vskip.6cm
\centerline{\bf REFERENCES}
\def\ref#1{\lbrack #1\rbrack}

\vskip.4cm
\par\noindent

\item{\ref{1}}
P. Ginsparg,
{\it Applied Conformal Field Theory},
Lectures given at Les Houches Summer School in Theoretical Physics, 
Les Houches, France, June--August, 1988,
Les Houches Summer School 1988 (1988), 1,
and references therein.

\item{\ref{2}}
P. Ginsparg,
in {\it Champs, Cordes et Ph\'enom\`enes Critiques},
E. Br\'ezin and J. Zinn--Justin eds, North Holland, Amsterdam (1989),
and references therein.

\item{\ref{3}}
J. Cardy,
in {\it Champs, Cordes et Ph\'enom\`enes Critiques},
E. Br\'ezin and J. Zinn--Justin eds, North Holland, Amsterdam (1989), 
and references therein.

\item{\ref{4}} 
H. Osborn,
{\it Implications of Conformal Invariance for Quantum Field 
Theories in $D>2$}, 
hep-th/9312176,
and references therein.

\item{\ref{5}} 
H. Osborn and A. Petkos,
{\it Implications of Conformal Invariance in Field Theories for 
General Dimensions},
Annals Phys. {\bf 231} (1994), 311, 
and references therein.

\item{\ref{6}} 
A. Cappelli and A. Coste 
{\it On the Stress Tensor of Conformal Field Theories in 
Higher Dimensions},
Nucl. Phys. {\bf B314} (1989), 707.

\item{\ref{7}} 
A. Sudbery,
{\it Quaternionic Analysis},
Math. Proc. Camb. Phil. Soc. {\bf 85} (1979), 199, 
and references therein.

\item{\ref{8}} 
R. S. Kulkarni,
{\it On the Principle of Uniformization}
J. Differential Geometry {\bf 13} (1978), 109.

\item{\ref{9}} 
R. S. Kulkarni,
{\it Conformal Structures and Moebius Structures}
in {\it Conformal Geometry}, a publication of the Max Planck Inst.
fuer Mathematik, R. S. Kulkarni and U. Pinkall eds., 
Aspects of Mathematics, Friedr. Vieweg \& Sohn Verlagsgesellschaft, 
Braunschweig, 
vol. {\bf E12} (1988), 1.

\item{\ref{10}} 
V. P. Nair and J. Schiff,
{\it A Kaehler Chern--Simons Theory and Quantization of Instanton Moduli 
Spaces},
Phys. Lett. {\bf 246} (1990), 423.
{\it Kaehler Chern--Simons Theory and Symmetries of Antiselfdual
Gauge Fields},
Nucl. Phys. {\bf B371} (1992), 329.

\item{\ref{11}} 
A. Losev, G. Moore, N. Nekrasov and S. Shatashvili, 
{\it Four--Dimensional Avatars of Two--Dimensional RCFT},
Talk given at Strings 95, Los Angeles, CA, March 1995, and at the
Trieste Conference on S Duality and Mirror Symmetry, Trieste, Italy, June
1995, 
Nucl. Phys. Proc. Suppl. {\bf 46} (1996), 130.
{\it Chiral Lagrangians, Anomalies, Supersymmetry and Holomorphy},
Nucl. Phys. {\bf B484} (1997), 196.

\item{\ref{12}}
H. B. Lawson Jr. and M.-L. Michelsohn
{\it Spin Geometry},
Princeton University Press,
Princeton (1989).

\item{\ref{13}}
J. G. Ratcliffe, 
{\it Foundations of Hyperbolic Manifolds},
Graduate Texts in Mathematics,
Springer Verlag, New York Berlin Heidelberg, 
{\bf 149} (1994).

\item{\ref{14}}
A. Besse, 
{\it Einstein Manifolds},
Ergebnisse der Mathematik und ihrer Grenzgebiete,
Springer Verlag, New York Berlin Heidelberg, 
3. Folge, Bd. {\bf 10} (1987).

\item{\ref{15}}
N. Birrel and P. Davies,
{\it Quantum Fields in Curved Space},
Cambridge University Press, 
Cambridge (1982).

\item{\ref{16}}
R. J. Riegert, 
{\it A Non--Local Action for the Trace Anomaly},
Phys. Lett. {\bf B134} (1984), 56.

\item{\ref{17}}
I. Antoniadis and E. Mottola,
{\it 4--d Quantum Gravity in the Conformal Sector}
Phys. Rev. {\bf D45} (1992), 2013.

\item{\ref{18}}
L. Brown and J. Cassidy,
{\it Stress Tensors and their Trace Anomalies in Conformally
Flat Space--Times},
Phys. Rev. {\bf D16} (1977), 1712,

\item{\ref{19}}
D. M. Capper and M. J. Duff,
{\it Trace Anomalies in Dimensional Regularization}
Nuovo Cimento {\bf 23A} (1974), 173.

\item{\ref{20}}
M. J. Duff,
{\it Observations on Conformal Anomalies},
Nucl. Phys. {\bf B125} (1977), 334.

\item{\ref{21}}
R. Zucchini, work in progress.

\bye